\documentclass[twocolumn,nofootinbib]{revtex4}
\usepackage[utf8]{inputenc}
\usepackage{amsmath}
\usepackage{graphicx}
\usepackage{caption}
\usepackage{amssymb}
\usepackage{color}
\usepackage{cancel}
\usepackage{framed}
\usepackage{comment}
\usepackage{mathtools}
\usepackage{hyperref}
\usepackage{chngcntr}
\usepackage[normalem]{ulem}
\usepackage[flushleft]{threeparttable}

\counterwithin{equation}{section}

\newcommand{\vev}[1]{\langle #1 \rangle}
\newcommand{\bra}[1]{\langle #1 |}
\newcommand{\ket}[1]{| #1 \rangle}
\newcommand{\nn}{\nonumber}

\newcommand{\dis}[1]{\displaystyle{#1}}

   \def\e{\epsilon} 
\def\d{\delta}
 \def\m{\mu} \def\n{\nu} \def\k{\kappa} \def\r{\rho}
\def\s{\sigma}   \def\l{\lambda}
\def\D{\Delta}  \def\O{\Omega}

\newcommand{\Acal}{{\mathcal A}} \newcommand{\Ocal}{{\mathcal O}}

 \newcommand{\Ical}{{\mathcal I}}

\newcommand*{\affaddr}[1]{#1} 
\newcommand*{\affmark}[1][*]{\textsuperscript{#1}}

\def\Cincy{\small{Department of Physics, University of Cincinnati, Cincinnati, Ohio 45221, USA}}
\def\Weizmann{\small{Department of Particle Physics and Astrophysics, Weizmann Institute of Science, Rehovot 761001, Israel}}

\begin{document}

\date{\today}
\title{\Large\bfseries On Approximation Methods in the Study of Boson Stars}

%

\author{
Joshua Eby\affmark[$\dag$], Madelyn Leembruggen\affmark[$\dag\dag$], Lauren Street\affmark[$\dag\dag$], Peter Suranyi\affmark[$\dag\dag$], \\ and L.C.R.Wijewardhana\affmark[$\dag\dag$] \\
{\it\affaddr{\affmark[$\dag$]\Weizmann}} \\
{\it\affaddr{\affmark[$\dag\dag$]\Cincy}}
}

\begin{abstract}

We analyze the accuracy of the variational method in computing physical quantities relevant for gravitationally bound Bose-Einstein condensates.  Using a variety of variational ans\"atze found in existing literature, we determine physical quantities and compare them to exact numerical solutions. We conclude that a ``linear+exponential" wavefunction proportional to $(1 + \xi)\exp(-\xi)$ (where $\xi$ is a dimensionless radial variable) is the best fit for attractive self-interactions along the stable branch of solutions, while for small particle number $N$ it is also the best fit for repulsive self-interactions.  For attractive self-interactions along the unstable branch, a single exponential is the best fit for small $N$, while a sech wavefunction fits better for large $N$. The Gaussian wavefunction ansatz, which is used often in the literature, is exceedingly poor across most of the parameter space, with the exception of repulsive interactions for large $N$.  We investigate a ``double exponential" ansatz with a free constant parameter, which is computationally efficient and can be optimized to fit the exact solutions in different limits. We show that the double exponential can be tuned to fit the sech ansatz, which is computationally slow.  We also show how to generalize the addition of free parameters in order to create more computationally efficient ans\"atze using the double exponential.  Determining the best ansatz, according to several comparison parameters, will be important for analytic descriptions of dynamical systems. Finally, we examine the underlying relativistic theory, and critically analyze the Thomas-Fermi approximation often used in the literature.
\end{abstract}

\maketitle

\section{Introduction}
The recent surge in the study of scalar field condensate dark matter (DM) is in part driven by the failure to detect individual Weakly Interacting Massive Particles (WIMPs) at the Large Hadron Collider, or in various direct detection experiments. There is another avenue for dark matter to manifest, as condensates of macroscopic size \cite{Kaup,RB,BreitGuptaZaks,CSW,SS,Friedberg,SS2,Liddel,Lee}. Electrically neutral boson particles, if they are a component of dark matter, can naturally form gravitationally bound Bose-Einstein condensate (BEC) bubbles below a critical temperature, due to quantum statistical effects. These entities are known either as \emph{oscillons} or \emph{boson stars}. It is quite interesting and important to understand the physical properties of such condensates to determine if they are a viable alternative to the more popular and widely investigated WIMPs. Thus it is important to develop analytical and numerical methods to study condensate formation, as well as their stability, evolution, and possible decay into component particles.

One widely studied example of scalar field DM forming BECs is the axion. Axion stars were considered first around 30 years ago, originally suggested to form from collapse of overdense miniclusters in the early universe \cite{HoganRees,KolbTkachev} (see also more recent simulations \cite{TkachevSim}). Since then, many properties of axion stars have been studied extensively; these include structural stability \cite{SikivieYang,ChavanisMR,ChavanisMR2,BB,BarrancoNS,ESVW,GHPW,Braaten,WilczekASt,Hertzberg} (including nonzero angular momentum \cite{Davidson,Sarkar,Hertzberg2}), the process of gravitational collapse \cite{Harko,ChavanisCollapse,ELSW,ELSW2,Levkov,Helfer,Michel}, and their decay through emission of relativistic particles \cite{ESW,MTY,BraatenEmission,EMSW,Hertzberg3,ELSWFDM}. There has recently been a significant amount of work regarding relativistic corrections more generally to the classical field description of axion stars \cite{GuthRelativistic,GRB,BraatenRelativistic,EMTWY}. Other authors have investigated possible connections to astrophysical radio sources \cite{Iwazaki,TkachevFRB}. In some scalar field models, boson stars can be extremely heavy and (if they are stable) could give rise to gravitational wave signatures \cite{GW1,GW2,GW3,GW4,GW5,GW6}. Clearly this is a field booming with new and interesting results. (For boson star reviews, see e.g. \cite{Jetzer,MielkeSchunck,Liebling}.)

Besides DM, there are other classes of problems in cosmology where condensate formation is important. Elementary Hermitian boson fields known as inflatons are postulated to drive cosmological inflation, the hypothesized early-time exponential expansion of the universe. In addition to this, a bosonic degree of freedom (termed quintessence) is posited to generate the energy density which causes the observed late time acceleration of the universe. Both inflatons and quintessence can fragment and form BEC bubbles in the same way as dark matter candidates discussed previously \cite{AminInf1,AminInf2,AminQuint}. If sufficiently long lived, these entities can play a crucial role during inflation and also at later stages of cosmological evolution.

Theoretical studies of Bose-Einstein condensation gained prominence in 1990’s, after the experimental
discovery of atomic BECs in systems of cold atoms \cite{Stoof,Arovas,PS}. Atomic condensates are described by the Gross-Pit\"aevskii (GP) equation, which is a form of the non-linear Schr\"odinger equation. The interatomic interactions could be attractive or repulsive, and the atoms could be placed on external potentials. Analytic solutions of the GP equation are difficult to obtain, and various approximation methods had to be employed. Static problems were more amenable to numerical studies, but dynamical questions like expansion, collapse, and decay of condensates required the employment of approximation methods. 

The variational method is one approximation technique widely utilized in studies of atomic condensates.  Recently, the variational method was also adopted for the study of gravitationally bound condensates of
bosons by Chavanis \cite{ChavanisMR}; he used a Gaussian ansatz to approximate the wavefunction and compared the results he obtained thus with numerical solutions of the GP equation \cite{ChavanisMR2}.  This comparison of ans\"atze to the numerical solutions is imperative for the study of dynamical problems that are much more difficult to solve numerically.  Some such dynamical problems that have been analyzed using the variational method are the collapse \cite{Harko,ChavanisCollapse,ELSW,ELSW2} and collisions \cite{Cotner,ELLSW} of BECs.  In subsequent years, numerous authors have presented various ans\"atze for both static and dynamical problems, either to improve numerical agreement or computational efficiency \cite{GHPW,Hertzberg,Hertzberg2,Hertzberg3}.\footnote{A third approach, in which the exact wavefunction can be computed using an analytical expansion, was developed in \cite{Kling1,Kling2}. This approach has the dual advantage of arbitrary precision and analyticity, though it is still more computationally intensive than the variational approach.}

To our knowledge, the relative efficacy of one ansatz compared to another has not previously been considered in a rigorous fashion. Some ans\"atze are simpler computationally, others are more complicated. Some match numerical results for boson star masses but do not reproduce as well the radius. As the literature on boson stars becomes more complex, it becomes increasingly important to have at hand a tractable analytic approach, and to understand the benefits and weaknesses of different choices of approximate wavefunctions. This is the basic goal of the current project.

In this work, we will provide background information for both the time-independent and time-dependent variational methods (Section \ref{sec:VarMethod}). The latter is crucial for understanding dynamical processes, like collapse and decay. We will then analyze various classes of approximate wavefunctions used in the literature, providing comparisons across different ans\"atze (Section \ref{sec:Ansatze}) as well as comparison to exact numerical results (Section \ref{sec:NumMethod}). In this discussion, we will also propose a new ansatz with a free parameter, which can be varied to give excellent agreement by whichever measure is relevant to a particular scenario. Finally, we compare the nonrelativistic formulation to the underlying relativistic theory in Section \ref{sec:RB}, and examine where the Thomas-Fermi approximation (often used in the literature) breaks down. We conclude in Section \ref{sec:Conclusions}.

We will use natural units throughout, where $\hbar = c = 1$.

\section{Variational Method} \label{sec:VarMethod}

\subsection{General Formulation}
The variational method~\cite{Ritz} for finding approximate analytic solutions to eigenvalue problems was utilized as an important calculational tool  in the early development of quantum theory. Its success in describing the ground state of Helium~\cite{Kellner} played an important role in establishing modern quantum mechanics as a viable theory.

As  explained in quantum mechanics texts, the variational method involves the extremization of  the expectation value of the Hamiltonian of the system $\bra{\Psi} H \ket{\Psi} $ with respect to variations of a class of candidate wave functions $\ket{\Psi}$. In the calculation, the norm $ \vev{\Psi|\Psi} $ is held fixed, and in the end one obtains approximate analytic solutions to the time-independent Schr\"odinger equation.  A corresponding method to obtain solutions to the time-dependent  Schr\"odinger equation is to extremize the action 
$$
S= \int dt\,d^3r  \left<\Psi(t)\left| i \frac{ \partial}{\partial t}-H \right|\Psi(t)\right>
$$
with respect to a class of time-dependent variational wave functions $ \left|\Psi(t)\right>$.  This time-dependent variational method, first introduced by Dirac in 1934 \cite{Dirac}, is not usually described in standard texts of quantum mechanics, but it has found wide use in fields like nuclear physics~\cite{Kerman}, chemistry~\cite{Langhoff}, and quantum field theory 
~\cite{Jackiw}.

In condensed matter discussions, the GP formalism is developed starting from an $N$-particle wave function where each particle is in the state $\phi(\vec{r})$. Following the derivations in \cite{PS}, the expectation value of the Hamiltonian 

\begin{align}
H = \sum_i \left(\frac{p_i^2}{2m} + V(\vec{r})\right) + U_0 \sum_{i<j} \delta^3\left(\vec{r}_i - \vec{r}_j\right)
\end{align}
in the state $\chi \left(\vec{r}_1, ... , \vec{r}_n\right) = \prod_i \phi(\vec{r}_i)$ yields the energy functional 
\begin{align}
E = N \int d^3 r  \left( \frac{\left| \nabla \phi \right|^2}{2m} + V(\vec{r}) \left|\phi \right|^2 + \frac{U_0}{2} \left(N-1 \right) \left|\phi \right|^4 \right)
\end{align}
where $V(\vec{r})$ is an external potential, $U_0$ represents some contact interaction among the scalars, and $\phi$ is normalized to unity. Defining $\psi = \sqrt{N} \phi(\vec{r})$ and taking $N\gg1$ yields,
\begin{align} \label{GPenergy}
E = \int d^3 r \left( \frac{\left|\nabla \psi \right|^2}{2m} + V(\vec{r}) \left|\psi \right|^2 + \frac{U_0}{2} \left|\psi \right|^4 \right).
\end{align}
Extremizing the above equation, subject to the normalization constraint
\begin{equation} \label{Norm}
\int d^3 r \left|\psi \right|^2 = N
\end{equation}
yields
\begin{align} \label{GPmu}
\mu \,\psi = -\frac{1}{2m} \nabla^2 \psi + c\,V(\vec{r})\, \psi + U_0 \left|\psi \right|^2 \psi,
\end{align}
where $\mu$, the Lagrange multiplier introduced to maintain the normalization, is defined as the chemical potential. The constant $c=1$ if $V(\vec{r})$ does not depend on $\psi^*$ (true for an external trap), but can take other values elsewhere (e.g. for condensates bound by self-gravity). By multiplying eq. (\ref{GPmu}) by $\psi^*$ and performing a volume integral, we get
\begin{align} \label{muint}
\mu N = \int d^3 r \left(\frac{\left|\nabla\psi\right|^2}{2m}+ c\, V(\vec{r})\,\left|\psi\right|^2 
		+ U_0 \left|\psi \right|^4\right).
\end{align}
Comparing eqs. (\ref{GPenergy}) and (\ref{muint}), we can conclude the chemical potential is not equal to the energy per particle given in eq. (\ref{GPenergy}) if the interaction strength $U_0 \neq 0$ or if the trapping potential $V(\vec{r})$ depends on $\psi^*$ (i.e. if $c\neq1$).

In the time-dependent case, the GP formalism starts with the action $S = \int dt L$, where 
\begin{align} \label{L(t)}
L = \int d^3 r \left( i \psi^* \dot{\psi} - i \psi \dot{\psi^*} - \frac{\left| \nabla \psi \right|^2}{2m} - V(\vec{r})\left| \psi \right|^2 - \frac{U_0}{2} \left|\psi \right|^4\right).
\end{align}
The Hamiltonian is 
\begin{align}
H &= p_{\psi} \dot{\psi} + p_{\psi^*} \dot{\psi^*} - L \\
&= \int d^3 r \left( \frac{\left|\nabla \psi \right| ^2}{2m} + V(\vec{r}) \left| \psi \right|^2 + \frac{U_0}{2} \left|\psi \right|^4 \right)
\end{align}
which is of course identical to the energy functional defined in equation (2.3). Variation of eq. (\ref{L(t)}) with respect to $\psi^*$ gives the Gross-Pit\"aevskii (GP) equation
\begin{equation} \label{GPtime}
 i\,\frac{\partial \psi}{\partial t} = -\frac{1}{2m} \nabla^2 \psi + c\,V(\vec{r})\, \psi + U_0 \left|\psi \right|^2 \psi,
\end{equation}
which reduces to the time-independent case of eq. (\ref{GPmu}) if the time dependence of $\psi$ is given by the factor $\exp(-i\,\mu\,t)$. This is also in agreement with the the identification of $\mu$ as the chemical potential.

The variational method has been applied to study time-independent as well as time-dependent solutions to the GP equation. As we will describe in Section \ref{sec:Ansatze}, in this case one describes the wavefunction by some ansatz $\psi(\vec{r})$, so that, for a given physical system, eq. (\ref{GPenergy}) can be extremized analytically. Of course, for many physical applications the solutions to the time-independent equations can be found numerically to an arbitrary precision; therefore the variational method offers qualitative understanding of the systems, but falls short of the numerical accuracy.

On the other hand, dynamical problems are more difficult to solve using numerical methods. Time-dependent solutions to the GP equation can be approximated using the variational method formulation of Pethick and Smith \cite{PS}, which we have described above. A good example of such an application is the collapse of BECs \cite{Harko,ChavanisCollapse,ELSW,ELSW2}. In this case, the variational parameter $\sigma$ is taken to be a function of time, $\sigma(t)$, while the wavefunction is multiplied by some phase.  This phase depends on the velocity field of the system, which for spherical collapse is taken to be in the radial direction and proportional to $r$.  It also depends on some parameter (analogous to the Hubble parameter in cosmology) which is related to $\sigma(t)$ by $H(t) = \dot{\sigma}(t)/\sigma(t)$. It is remarkable that using a simple analytic description of this type, one can assess many of the relevant features of BEC collapse. This illustrates the power of the time-dependent variational formalism.

\subsection{Application to Boson Stars}
Dynamics of boson stars are described by the Klein-Gordon equation in the presence of self-gravity. This action,
\begin{align}
S = \int \sqrt{-g} \,dt \,d^3 r \left[\frac{1}{2} g_{\mu \nu} \left(\partial^{\mu} \phi\right) \left(\partial^{\nu} \phi \right) - \frac{m^2}{2} \phi^2 - \frac{\lambda}{4!} \phi^4\right]
\end{align}
in the nonrelativistic limit and with the replacement ${\phi = \frac{1}{\sqrt{2m}} \left(e^{-i m t} \psi + h.c.\right)}$ gives
\begin{align} \label{Action}
 S = \int dt\, d^3r \Big[&i \psi \frac{ \partial \psi^*}{\partial t}-i \psi^* \frac{\partial \psi}{\partial t} + \frac{|\nabla \psi|^2}{2m}  \nn \\
	&+ \frac{1}{2} V_{g}\,|\psi|^2
	+  \frac{\l}{16\,m^2}\left|\psi\right|^4 \Big].
\end{align}
The gravitational potential
\begin{equation} \label{Poisson}
 \nabla^2 V_g = 4\pi\,G\,m^2\,|\psi|^2
 \end{equation}
originates from the nonrelativistic limit of the Einstein field equations with $G=M_P{}^{-2}$ Newton's gravitational constant. (See Section \ref{sec:RB} for a more thorough description of the nonrelativistic limit.) In this work, we will consider the case of a quartic self-interaction, where self-coupling $\lambda$ can be positive (giving rise to a repulsive interaction) or negative (attractive interaction). Variation of eq. (\ref{Action}) with respect to $\psi^*$ yields a GP-type equation comparable to eq. (\ref{GPtime}).
\begin{equation} \label{GPtime2}
 i \frac{ \partial \psi}{\partial t} =  -  \frac{1}{2m}\nabla^2 \psi
	+  V_{g}\,\psi 
	+  \frac{\l}{8\,m^2}\left|\psi\right|^2 \psi.
\end{equation}
Finding analytic solutions to the above non-linear equations (\ref{Poisson}) and (\ref{GPtime2}), known as the Gross-Pit\"aevskii+Poisson (GPP) system, is a challenging task. In order to find an approximate analytic solution  one could extremize the action $S$ within a class of variational wave functions as described previously in condensed matter/atomic physics applications. 

Assuming a harmonic time dependence for the wavefunction, the chemical potential of eq. (\ref{muint}) is given by
\begin{equation} \label{muint2}
 \mu\,N =  \int d^3 r \left(\frac{\left|\nabla\psi\right|^2}{2m}+ V_{g}\,\left|\psi\right|^2 
		+ \frac{\l}{8\,m^2} \left|\psi \right|^4\right).
\end{equation}
while the energy of eq. (\ref{GPenergy}) is 
\begin{equation} \label{Energy}
  E = \int d^3r \left(\frac{|\nabla \psi|^2}{2m} 
	+ \frac{1}{2} V_{g}\,|\psi|^2 
	+  \frac{\l}{16\,m^2}\left|\psi\right|^4 \right).
\end{equation}
The latter can be extremized by assuming some $r$ dependence for $\psi$, giving rise to an approximate bound state solution. We present a number of prevalent choices for such an ansatz in the next section. We will focus here on the time-independent case, because it will allow us to analyze the radial dependence of the wavefunction; any relevant time-dependent factors will depend on the particular application one considers (e.g. collapse) and can be added on later.

In order to begin, we will need to impose some conditions on the classes of wavefunction ans\"atze we consider. Assuming that the wave function $\psi(r) $ is differentiable to all orders at $r=0$ and the gravitational potential $V_g$ has a Laurent expansion in $r$, the GPP equations (\ref{Poisson}) and (\ref{GPtime2}) imply that the gravitational potential is an analytic function in the variable $ r^2 $ at the origin; consequently, the first derivative of the wave function $\psi(r) $ vanishes there.  At large $r$, the interaction term in the GP equation becomes negligibly small, and the GP equation reduces to the linear  Schr\"odinger equation for a particle moving in the gravitational potential $V_g$. The potential at large $r$ takes the $G M/ r $ Newtonian form, and the wave functions asymptotically are nearly hydrogen-like, with a $\exp(-kr) $ behavior (though see below). We impose these boundary conditions when numerically solving the GPP equations. A successful variational ansatz should also exhibit similar behavior. 

Actually, the behavior of the wavefunction at large $r$ is not exactly exponential, but can be calculated in a straightforward way. First, at large $r$ we throw away the terms in the equation of motion that are higher order in the fields than the gravity term; this gives eq. (\ref{GPtime2}) with $\lambda=0$ and $V_g = -G M m / r$. This equation has an exact solution in terms of hypergeometric functions, but the leading order solution is proportional to
\begin{equation} \label{asymptotic}
 \psi \propto e^{-\sqrt{-2\,\mu\,m}\,r} 
                             r^{\frac{G\,m^2\,M}{\sqrt{-2\,\mu\,m}} - 1}.
\end{equation}
This is in agreement with the results of \cite{Kling1,Kling2}.

Compare and contrast this case to that of atomic Bose condensates trapped by external harmonic oscillator potentials. In the latter situation, the GP wave function at large $r$, when interaction terms are negligible, should approach Gaussian form,  making such functions ideal candidates for variational  ans\"atze. This makes clear the fact that the optimal ansatz for the wavefunction should depend on the potential.

\section{Ans\"atze for Boson Star Wavefunctions} \label{sec:Ansatze}
\subsection{Generalized Ansatz}

In \cite{ELSW,ELSW2} we performed an analysis using a general ansatz for the wavefunction of an axion star, under only the (weak) assumptions that the wavefunction is finite at the origin, spherically symmetric, and decreases monotonically with the radial coordinate $r$. Here, we apply the same method to a general boson star with a 4-point coupling $\l$. 

The starting point is the rescaling of the macroscopic quantities
\begin{equation} \label{scaling}
 \s = \sqrt{|\l|}\frac{M_P}{m^2}\,\rho \qquad N = \frac{M_P}{m\,\sqrt{|\l|}}\,n.
\end{equation}
where $\s$ is a variational distance parameter explained in detail in the next section. Then a general ansatz for a boson star can be written in the form
\begin{equation}\label{ansatz}
\psi(r)= w\, F\left(\frac{r}{\s}\right)\equiv \psi(0)\,F(\xi),
\end{equation}
where at fixed $\xi$ the function $F(\xi)$ is independent of $\rho$, $n$, and $\l$. Because we can fix $F(0)=1$ without loss of generality, we are able to identify $w = \psi(0)$, the central value of the wavefunction. Substituting the ansatz into the normalization condition (\ref{Norm}) gives the central value as
\begin{equation}
\psi(0) = \frac{m^{5/2}}{M_P\,|\l|}\,\sqrt{\frac{n}{\rho^3\,C_2}},
\end{equation}
where we introduced the notation
\begin{equation}
C_k = 4\,\pi\int_0^\infty d\xi\,\xi^2 \,F(\xi)^k.
\end{equation}

Using eq. (\ref{Energy}) and the general ansatz of eq. (\ref{ansatz}), we obtain for the energy functional
\begin{equation}\label{binding}
 \frac{E(\rho)}{m\,N} = \frac{m^2}{M_P{}^2\,|\l|}\left(\frac{D_2}{2\,C_2}\frac{1}{\rho^2}-\frac{B_4}{2\,C_2{}^2}\,\frac{n}{\rho}
	+ \text{sgn}(\l) \frac{C_4}{16\,C_2{}^2}\frac{n}{\rho^3}\right),
 \end{equation}
and, using eq. (\ref{muint2}), we obtain for the chemical potential
\begin{equation}\label{Chemical}
 \frac{\mu(\r)}{m} = \frac{m^2}{M_P{}^2\,|\l|}\left(\frac{D_2}{2\,C_2}\frac{1}{\rho^2}-\frac{B_4}{C_2{}^2}\,\frac{n}{\rho}
 			+ \text{sgn}(\l) \frac{C_4}{8\,C_2{}^2}\frac{n}{\rho^3}\right).
 \end{equation}
For simplicity, we have defined the dimensionless parameters
\begin{align}
D_2&=4\,\pi\int_0^\infty d\xi\,\xi^2 \,F'(\xi)^2,\\
B_4&=32\,\pi^2\int_0^\infty d\xi\,\xi \,F(\xi)^2\int_0^\xi d\eta\, \eta^2\,F(\eta)^2.
\end{align}
Recall once more that $\l>0$ ($\l<0$) will correspond to repulsive (attractive) self-interactions.

Because the GP equation can be derived from the variation of the total energy, a solution will be a stationary point of (\ref{Energy}). Given an ansatz for the wavefunction, we approximate this exact solution by minimizing eq. (\ref{binding}) with respect to $\rho$. This procedure gives the dilute boson star radius
\begin{equation} \label{rhodilute}
 \rho_d = \frac{C_2\,D_2}{B_4\,n}\left[1 + \sqrt{1 + \text{sgn}(\l) \frac{n^2}{\bar{n}^2}}\right].
\end{equation}
where $\bar{n}$ is an important scaled particle number given by 
\begin{equation} \label{nCrit}
 \bar{n} = \sqrt{\frac{8}{3}}\,\frac{C_2\,D_2}{\sqrt{B_4\,C_4}}.
\end{equation}
For $\l<0$, $\bar{n}$ determines the mass of the maximum stable configuration, and so we will denote it by $\bar{n}=n_c$. For $\l>0$, there is no maximum mass in the nonrelativistic limit, so we will instead use the notation $\bar{n}=n_*$; in this case, self-interactions become increasingly important for $n\sim n_*$, eventually approaching the region in which the Thomas-Fermi approximation is relevant. We will discuss these points in detail below.

Eq. (\ref{rhodilute}) is a stable minimum of the energy classically and a metastable solution in quantum theory. There exists another root of the energy at a radius of
\begin{equation} \label{rhounstable}
 \rho_{u} = \frac{C_2\,D_2}{B_4\,n}\left[1 - \sqrt{1 + \text{sgn}(\l) \frac{n^2}{\bar{n}^2}}\right],
\end{equation}
which we will discuss later. In the case of attractive self-interactions, $\r_u>0$ is an unstable maximum of the energy; for repulsive interactions, $\r_u < 0$ is an unphysical configuration.

Above, we have used the formulation of our previous works \cite{ELSW,ELSW2}, in which the dimensionful parameters are scaled out; then the numerical constants $B_4$, $C_k$, and $D_2$ are dimensionless and depend only on the shape of the ansatz being employed. Other authors, e.g. \cite{ChavanisMR,EKNW,ESVW2}, use a different formulation; for ease of comparison we provide the expressions for translating between the two in Appendix \ref{AppB}.

This formulation has a number of useful applications.  For example, in the limit of no self-interactions ($\lambda\to0$)\footnote{Of course, in the $\lambda\to0$ limit the scaling of eq. (\ref{scaling}) is not appropriate. In that case one should use $\s = \rho/m$ and $N = M_P{}^2\,n/m^2$, which gives rise to the analogous equations for $E$ and $\mu$ without the interaction term. In the end, one finds the first two terms in eqs. (\ref{binding}) and (\ref{Chemical}) do not change, which is all that is needed here.}, there is a simple relation between the energy and chemical potential:
\begin{align} \label{muoverE_NI}
 \frac{N\,\m(\r_d)}{E(\r_d)} &= \frac{\frac{D_2}{2\,C_2}\frac{1}{\rho_d^2}-\frac{B_4}{C_2{}^2}\,\frac{n}{\rho_d}}
 					{\frac{D_2}{2\,C_2}\frac{1}{\rho_d^2}-\frac{B_4}{2C_2{}^2}\,\frac{n}{\rho_d}} \nn \\
			&= \frac{-3/8}{-1/8} \nn \\
			&= 3,
\end{align}
which was derived using other methods in \cite{Membrado,ChavanisMR}.

For attractive interactions, we of course recover the standard result that the local minimum and maximum of the energy become equal at $n=n_c$.
For $n>n_c$, no stable solutions exist.
The value $n=n_c$ also corresponds to a critical minimum radius,
\begin{equation} \label{rhoCrit}
 \rho_c = \sqrt{\frac{3 C_4}{8B_4}}.
\end{equation}
It is useful to define a parameter 
\begin{align}
\delta = \sqrt{1 - \left(\frac{N}{N_c}\right)^2} = \sqrt{1 - \left(\frac{n}{n_c}\right)^2}
\end{align}
which parameterizes the closeness of $N$ to $N_c$. Then substituting $\rho_d$ from eq. (\ref{rhodilute}) into eqs. (\ref{binding}) and (\ref{Chemical}),  the energy and chemical potential at the minimum for attractive self-interactions can be written in the form
\begin{align}
\frac{E(\rho_d)}{m \,N} &= -\frac{1-\delta}{1+\delta}(1+2\,\delta) \frac{m^2}{M_P{}^2 |\lambda|} \frac{4D_2}{9C_2}\frac{B_4}{C_4}\nn\\
\frac{\mu(\rho_d)}{m} &= -\frac{1-\delta}{1+\delta}(5+4\,\delta) \frac{m^2}{M_P{}^2 |\lambda|} \frac{4D_2}{9C_2}\frac{B_4}{C_4}.
\end{align}
This implies a simple relationship between the energy and the chemical potential that is independent of the choice of ansatz,
\begin{equation}
\frac{N\,\mu(\rho_d)}{E(\rho_d)}=\frac{5+4\,\delta}{1+2\,\delta},
\end{equation}
which is an analogue of the non-interacting result in eq. (\ref{muoverE_NI}), but applied to attractive self-interactions.
We can see that when $\d\to0$ ($N\to N_c$), the ratio goes exactly to $5$. Even though the derivation assumed some ansatz for the variational approach, the result does not depend on what form the wavefunction takes, and so it holds even in the exact case.

In the case of repulsive self-interactions, there also exists a critical mass, though it is not at $n=n_*$ as defined above; it arises due to relativistic effects which we do not consider here \cite{CSW}. Relativistic effects can be taken into account using the Ruffini-Bonazzola (RB) formalism for analyzing boson stars \cite{RB}. One could in principle formulate a variational method which approximated the relativistic equations of motion, in which case these effects would become apparent. We leave such an analysis for future work. Of course, at weak gravity and small binding energy, the RB equations of motion reduce to the GPP system, as we will describe in Section \ref{sec:RB}.

\subsection{Non-Compact Ans\"atze}

It is well-known by direct solution of eqs. (\ref{Poisson}) and (\ref{GPtime2}) that the wavefunction of a boson star does not typically have compact support; it is nonzero for all $r\geq 0$ though decreases extremely fast at large $r$. The standard definition for the ``size" of a boson star is $R_{99}$, the radius inside which $0.99$ of the mass is contained. A number of non-compact ans\"atze have appeared in the literature to approximate the exact solution; a few of the most popular ones are
\begin{widetext}
\begin{align} \label{eq:ansatze}
 \psi_A(r) =
 \begin{dcases}
  \sqrt{\frac{N}{\pi^{3/2}\,\s^{3}}}\,e^{-r^2/2\,\s^2} & \text{(Gaussian (G) \cite{ChavanisMR,ChavanisCollapse,ELSW,ELSW2,ELLSW,Hertzberg2,Hertzberg3})} \\
  \sqrt{\frac{N}{\pi\,\s^3}}\,e^{-r/\s} & \text{(Exponential (E) \cite{GHPW,Hertzberg})} \\
  \sqrt{\frac{N}{7\,\pi\,\s^3}}\left(1 + \frac{r}{\s}\right)\,e^{-r/\s} & \text{(Linear + Exponential (LE) \cite{Hertzberg})} \\
  \sqrt{\frac{3\,N}{\pi^3\,\s^3}}\,\text{sech}\left(\frac{r}{\s}\right) & \text{(Sech (S) \cite{Hertzberg,Hertzberg3})} \\
  \sqrt{\frac{N}{\pi\,\s^3}}\left(1 + \frac{1}{a^5} - \frac{16}{a(1+a)^3}\right)^{-1/2}\,
  			\left(e^{-r/\s} - \frac{1}{a}e^{-a\,r/\s}\right) & \text{(Double Exponential (DE$_a$))}
 \end{dcases}
\end{align}
\end{widetext}
On the right we give the long name (e.g. ``Gaussian") and short abbreviation (e.g. ``G") for each ansatz, and cite a collection of previous works which utilize them in the study of boson stars. We will use the notation that, for example, $\psi_G(r)$ is the Gaussian wavefunction, $\psi_{LE}(r)$ is the linear+exponential wavefunction, etc. The first four functions listed here are popular choices in the literature; the last one (the double exponential) is a proposal of ours with some constant parameter $a$ which can be fixed by matching to the exact solutions. We will show in this work that the double exponential can be optimized for a given numerical result. 

Note the appearance of $\s$ in each of the ans\"atze. While this parameter has units of distance, it should not be confused with the radius of a boson star. For each ansatz, the parameter $\s_d$ of the solution is related by some constant factor to the radius of the boson star $R_{99}$; we will label this by a real number $\kappa$, i.e.
\begin{equation}
 \kappa\ = \frac{R_{99}}{\sigma_d},
\end{equation}
where both $\sigma_d$ and $\kappa$ depend on the ansatz under consideration.
Other reasonable distance scales in a calculation like this one include the expectation values 
\begin{align}
 \vev{r} &\equiv \frac{1}{N}\int d^3r\,r\,|\psi(r)|^2 \nn \\
  \vev{r^2} &\equiv \frac{1}{N}\int d^3r\,r^2\,|\psi(r)|^2,
\end{align}
which will be useful for comparing to exact solutions later. Because the translation to a physical length is different for a different ansatz, the parameter $\s$ means something different depending on the ansatz in which it is employed; said a different way, $\s$ itself is unphysical and should not be compared across ans\"atze.

\begin{table*}
\centering
\begin{tabular}{| c || c | c | c | c | c | c | c |}
\hline
& G
& E 
& LE 
& S 
& DE$_2$
& DE$_{3\pi/8}$ \\
\hline \hline
 $F(\xi)$ & $\dis{e^{-\xi^2/2}}$
 		& $\dis{e^{-\xi}}$ 
		& $\dis{\left(1 + \xi \right)\,e^{-\xi}}$
		& $\dis{\text{sech}\left(\xi\right)}$
		& $\dis{2 e^{-\xi} - e^{-2\xi}}$
		& $\dis{\approx{6.6e^{-\xi} - 5.6e^{-1.2\xi}}}$
		\\ \hline
 $B_4$ &  $\dis{\sqrt{2\pi}\,\pi^2}$
 		& $\dis{\frac{5\,\pi^2}{8}}$
		& $\dis{\frac{5373\,\pi^2}{256}}$
		& $\dis{8\pi\left(2\pi\zeta(3) - \frac{\pi^3}{6}\right)}$
		& $\dis{\approx{47}}$
		& $\dis{\approx{61}}$
		\\ \hline
 $C_2$ &  $\dis{\pi^{3/2}}$
 		& $\dis{\pi}$
		& $\dis{7\pi}$
		& $\dis{\frac{\pi^3}{3}}$
		& $\dis{\approx{9.2}}$
		& $\dis{\approx{17}}$
		\\ \hline
 $C_4$ &  $\dis{\left(\frac{\pi}{2}\right)^{3/2}}$
 		& $\dis{\frac{\pi}{8}}$
		& $\dis{\frac{437\pi}{256}}$
		& $\dis{\frac{2\pi}{9}\left(\pi^2-6\right)}$
		& $\dis{\approx{2.1}}$
		& $\dis{\approx{4.2}}$
		\\ \hline
 $D_2$ &  $\dis{\frac{3\,\pi^{3/2}}{2}}$
 		& $\dis{\pi}$
		& $\dis{3\pi}$
		& $\dis{\frac{\pi(12+\pi^2)}{9}}$
		& $\dis{\approx{6.7}}$
		& $\dis{\approx{8.7}}$
		\\ \hline
 $\bar{n}$ & $\dis{2\pi\,\sqrt{3}}$
 		& $\dis{\sqrt{\frac{512\pi}{15}}}$
		& $\dis{\frac{3584}{3}\sqrt{\frac{2\pi}{86963}}}$
		& $\dis{\frac{2}{3}\frac{\left(\frac{1}{6} + \frac{2}{\pi^2}\right)\pi^{9/2}}{\sqrt{[\pi^2 - 6][12\zeta(3) - \pi^2]}}}$
		& $\dis{\approx{10.1255}}$
		&  $\dis{\approx{10.1518}}$
		\\ \hline
 $\r_c$  & $\dis{\sqrt{\frac{3}{32\pi}}}$
 		& $\dis{\sqrt{\frac{3}{40\pi}}}$
		& $\dis{\frac{1}{6}\sqrt{\frac{437}{398\pi}}}$
		& $\dis{\frac{\sqrt{[\pi^2 - 6][12\zeta(3) - \pi^2]}}{\sqrt{\pi}[48\zeta(3) - 4\pi^2]}}$
		& $\dis{\approx{0.13}}$
		& $\dis{\approx{0.11}}$
		\\ \hline
 $F'(0)$ & 0
 		& $<0$
		& 0
		& 0
		& 0
		& 0
		\\ \hline
 $\dis{\k}$ & $\dis{2.8}$
  		& $\dis{4.2}$ 
		& $\dis{5.4}$
		& $\dis{3.3}$ 
		& $\dis{3.4}$
		& $\dis{4.7}$
		\\ \hline
  \hline
\end{tabular}
\caption{Table of various parameters for the ans\"atze under consideration. Exact solutions for the double-exponential are given in Appendix \ref{AppC} for arbitrary values of the parameter $a$.}
\label{tab1}
\end{table*}

\subsection{Compact Ans\"atze} \label{compact_ansatze}

It is sometimes advantageous to employ a compact function to approximate the wavefunction. An example of such a case is the Thomas-Fermi (TF) limit of repulsive interactions, where one neglects the kinetic energy term, and the resulting wavefunction is very close to having an exact radius $R$. We also found a compact ansatz advantageous when discussing collapse of an axion star through collisions with astrophysical sources \cite{ELLSW}. Here, we include a few compact ans\"atze for completeness:
\begin{align}
 \psi_A(r) =
 \begin{dcases}
  \sqrt{\frac{4\,\pi\,N}{(2\pi^2 - 15)\,R^3}}\,\cos^2\left(\frac{\pi\,r}{2\,R}\right) & \text{(Cos$^2$ \cite{ELSW,ELLSW})} \\
\sqrt{\frac{\pi N}{4 R^3} \frac{\sin\left(\pi r/ R\right)}{\left(\pi r/R\right)}} & \text{(TF \cite{BH_TF})}
 \end{dcases}
\end{align}
However, we do not include them in our analysis, partly because it is somewhat problematic to compare $R_{99}$ for non-compact ans\"atze to the exact radius for compact ans\"atze. It should also be noted that one must require a compact ansatz to vanish above $r = R$.  The TF ansatz given above is the exact solution to the GPP equations in the Thomas-Fermi limit \cite{BH_TF}. However, given that the solution has an infinite derivative at $r = R$, we do not recommend its use as an ansatz.

\section{Comparison with Numerical Calculation} \label{sec:NumMethod}

\subsection{Numerical Algorithm}

To investigate the efficacy of these ans\"atze further, we have to solve the full equations of motion (\ref{Poisson}) and (\ref{GPtime2}) in the stationary limit. We employ the following scaling to dimensionless (``tilde-d") variables:
\begin{align} \label{eq:numscale}
 \psi &= \frac{m^{5/2}}{M_P\,|\l|}\,\tilde{\psi} \nn \\
 V_g &= \m + \frac{m^3}{M_P{}^2\,|\l|}\,\tilde{V} \nn \\
 r &= \frac{\sqrt{|\l|}\,M_P}{m^2}\,\tilde{r}
\end{align}
Writing the equations of motion in terms of the rescaled quantities, we solve the resulting system numerically:
\begin{align}
 \tilde{\nabla}^2 \tilde{\psi} &= 2\,\tilde V\,\tilde{\psi} + \frac{\text{sgn}(\l)}{4} \tilde{\psi}^3, \nn \\
 \tilde{\nabla}^2 \tilde{V} &= 4\pi\,\tilde{\psi}^2.
\end{align}
We employ a shooting method to determine correct boundary conditions $\tilde{\psi}\to 0$ and $\tilde{V}\to constant$ as $\tilde{r}\to\infty$. In practice, in the numerical routine $\tilde\psi$ and $\tilde{V}$ converge up to some finite radius $\tilde{r}_0$, which can be as large as the precision of the calculation requires. 
After solving for $\tilde\psi$ and $\tilde V$, one can calculate the dimensionless macroscopic quantities using
\begin{align}
 n &= \int_0^{\tilde{r}_0} d\tilde{r}\,4\pi\,\tilde r^2\,\tilde{\psi}^2 \nn \\
 .99\,n &= \int_0^{\tilde{R}_{99}} d\tilde{r}\,4\pi\,\tilde{r}^2\,\tilde{\psi}^2 \nn \\
 \tilde\mu &= - \lim_{\tilde{r}\to\tilde{r}_0} \tilde{V},
\end{align}
which determine the rescaled number, radius, and chemical potential (respectively). To convert back into standard physical units, we use
\begin{align} \label{macro}
 M &= \frac{M_P}{\sqrt{|\l|}}\,n , \nn \\
 R_{99} &= \frac{M_P}{m^2}\sqrt{|\l|}\,\tilde{R}_{99} , \nn \\
 \mu &= \frac{1}{|\l|}\frac{m^3}{M_P{}^2}\,\tilde\mu.
\end{align}

Before describing the solutions, note that the GPP energy can be written in terms of tilde-d quantities,
\begin{align} \label{DimEnergy}
 E(\tilde{\psi}) = \frac{m^2}{M_P\,|\l|^{3/2}} \int d^3\tilde r \Big[\frac{1}{2}  \tilde\nabla & \tilde\psi|^2 
	+ \frac{1}{2} \left(\tilde V + \tilde\mu\right)\,\tilde\psi^2 \nn \\
	&+  \frac{\text{sgn}(\l)}{16}\tilde\psi^4 \Big].
\end{align}
Using $M = m\,N$ and eq. (\ref{macro}) for $\tilde N$, we can write the energy per particle as
\begin{align}
 \frac{E(\tilde{\psi})}{m\,N} = \frac{m^2}{M_P{}^2\,|\l|} \frac{1}{n}\int d^3\tilde r\,\Big[
 		&\frac{1}{2}|\tilde\nabla \tilde\psi|^2
			+ \frac{1}{2} \left(\tilde V + \tilde\mu\right)\,\tilde\psi^2 \nn \\
			&+  \frac{\text{sgn}(\l)}{16}\tilde\psi^4\Big].
\end{align}
A similar procedure for the chemical potential gives
\begin{align}
 N\,\mu(\tilde\psi) = \frac{m^2}{M_P\,|\l|^{3/2}} \int d^3\tilde r \Big[\frac{1}{2}|\tilde\nabla& \tilde\psi|^2 
	+ \left(\tilde V + \tilde\mu\right)\,\tilde\psi^2 \nn \\
	&+ \,\frac{\text{sgn}(\l)}{8}\,\tilde\psi^4 \Big],
\end{align}
so that, after dividing by $N$ and $m$, we get
\begin{align}
  \frac{\mu(\tilde{\psi})}{m} = \frac{m^2}{M_P{}^2\,|\l|} \frac{1}{n}\int d^3\tilde r\,\Big[
 		&\frac{1}{2}|\tilde\nabla \tilde\psi|^2 
			+ \left(\tilde V + \tilde\mu\right)\,\tilde\psi^2 \nn \\
			&+  \frac{\text{sgn}(\l)}{8}\tilde\psi^4 \Big].
\end{align}
Of course, because $\mu$ appears on both the LHS and RHS, this equation can be used as a consistency check on the numerical solutions. We have verified for every numerical solution that the eigenvalue $\mu$ satisfies this constraint.

\begin{figure*}
 \includegraphics[scale=.53]{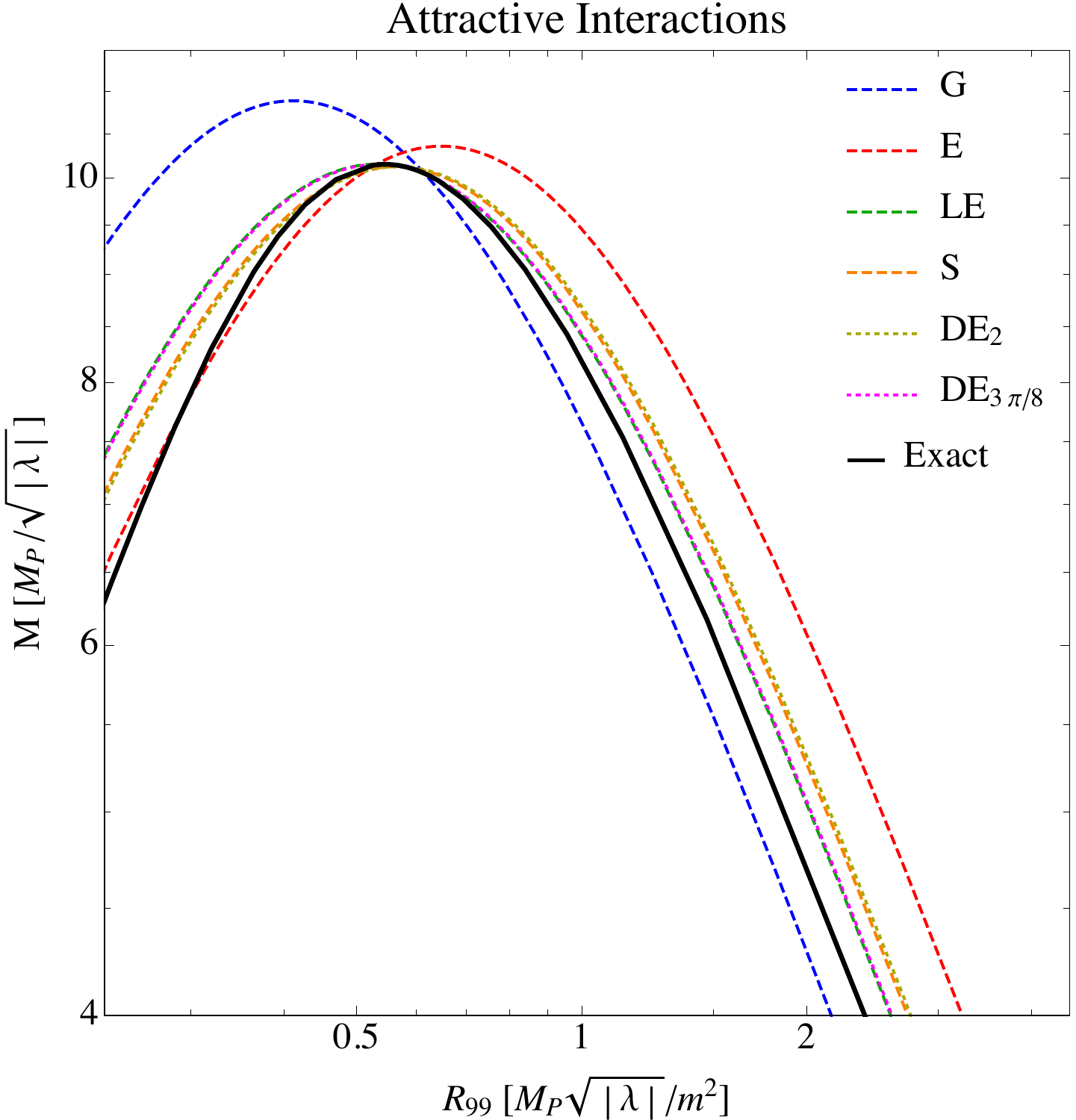}
 \includegraphics[scale=.55]{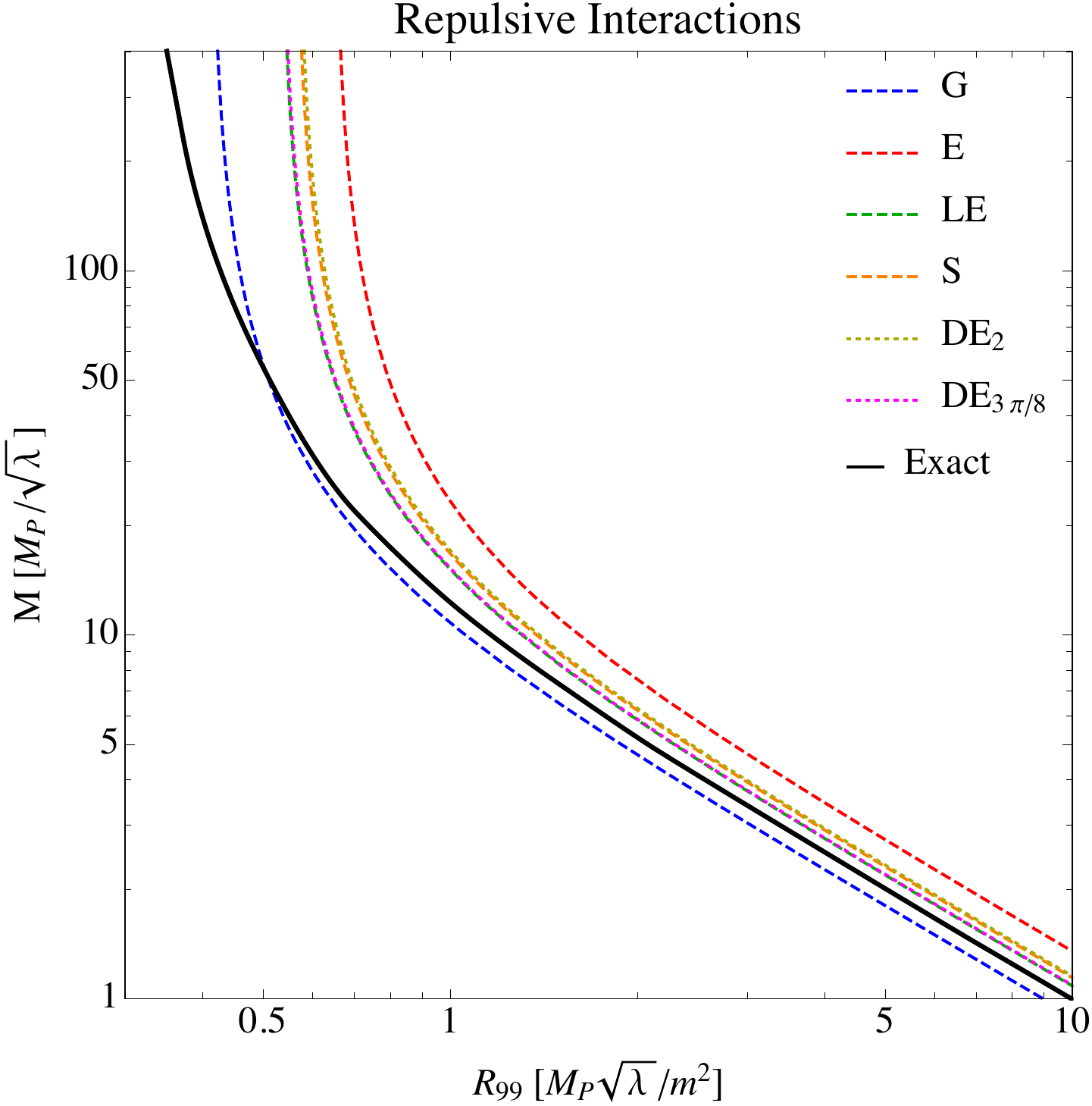}
 \caption{The rescaled mass $M$ and radius $R_{99}$ for boson stars. The black lines denote the exact numerical solutions, whereas the ans\"atze are color-coded in the legend. The left panel is the attractive interaction case, where the maximum mass is calculable in a nonrelativistic method; the right panel is the repulsive interaction case.}
 \label{fig:MassRadius}
\end{figure*}

The important physical quantities describing a boson star are its mass and radius. In Figure \ref{fig:MassRadius}, we show the exact numerical relationship between these quantities, along with the result from various ans\"atze.  For attractive interactions, we observe that $\psi_E$ tends to be a good fit for small $R_{99}$ (on the unstable branch of solutions) but a poor fit for large $R_{99}$ (the stable branch).  It is also interesting to note that $\psi_G$ is used quite often in the literature, but is a poor fit for small $R_{99}$ and badly approximates the position of the maximum mass.  The double exponential ansatz has a free parameter $a$, and we find that we can match the value of $n_c$ almost exactly by taking the value $a = 3\pi/8$. 

One interesting feature that is apparent in the right panel of Figure \ref{fig:MassRadius} is that, in the case of repulsive interactions, the radius at large $M$ approaches a constant. This is the limit of the Thomas-Fermi approximation, which will be discussed in Section \ref{sec:RB}. When using the variational approach, the radius $R_{TF}$ to which a given ansatz approaches at large $M$ in the repulsive case is equal to $R_c$, the radius at which one obtains the maximum mass in the attractive case. (This is clear by examination of eq. (\ref{rhodilute}) as well.) This relationship between these two radii in very different limits is a consequence of the fact that we use the same form for the ansatz in both the repulsive and attractive interaction cases. As a result, it does not hold true for the numerical solutions.

It is interesting that the result for the double exponential ansatz almost exactly approximate the sech and linear+exponential ans\"atze for the choices $a=2$ and $a=3\pi/8$, respectively. We will discuss this in more detail in the following section.

\subsection{Fitting the Double Exponential} \label{sec:DE}
Every variational calculation, regardless of the ansatz, can be improved by the addition of an extra parameter. The double exponential ansatz proposed in this paper has an extra variational parameter that can be tuned to fit the numerical solution and is computationally efficient.  The extra variational parameter allows one to choose how to fit the ansatz to the numerical solution.  For example, one can choose to fit the ansatz and numerical wavefunctions, or to fit various expectation values. In the comparisons above, we found that the choice $a = 3\pi/8$ reproduced the exact value of the maximum mass. This choice also resulted in an ansatz, denoted $\psi_{DE_{3\pi/8}}$, that was in very good agreement with $\psi_{LE}$.  We also considered the variational parameter $a = 2$, which resulted in an ansatz $\psi_{DE_2}$ that was in good agreement with $\psi_S$. By varying the parameter $a$, we found we could optimize agreement to the numerical solutions on both branches of attractive solutions, as well as for repulsive interactions.

As noted previously, $\psi_{DE_a}$ can be tuned to fit $\psi_{LE}$ and $\psi_{S}$.  We can explain this by looking at the double exponential in different limits of $a$. First of all, it is clear that in the limit $a\gg 1$, we have $\psi_{DE_a} \to \psi_E$, the ordinary exponential ansatz. On the other hand, in the limit $a\to 1$, 
\begin{align}
 \psi_{DE_a} &= f(a) \left(1 - \frac{1}{a}e^{-(a-1)r/\s}\right)e^{-r/\s} \nn \\
 		&\approx f(a) \left(1 - \frac{1}{a}\left[1 - (a-1)\frac{r}{\s}\right]\right)e^{-r/\s} \nn \\
		&= f(a) \left(1 - \frac{1}{a}\right) \left[1 + \frac{r}{\s}\right] e^{-r/\s},
\end{align}
where $f(a)$ is the original prefactor in eq. (\ref{eq:ansatze}). Because the prefactor in any ansatz is finite and equal to $\psi(0)$ by definition, we find that $\psi_{DE_a} \to \psi_{LE}$ in the limit $a\to1$. In this sense, the double exponential ansatz interpolates between $\psi_E$ and $\psi_{LE}$ as $a$ varies between $1$ and $\gg1$.

Slightly more perplexing is the sech ansatz, which seems to be reproduced approximately when $a=2$.  The reason for that is that the second derivatives at $r=0$ coincide for those ans\"atze.  The behavior near the origin is the most important  for all integrals over the wave functions, because the largest contributions to all integrals come from that region. This explains this coincidence and further illustrates the versatility of $\psi_{DE_a}$.

The double exponential can be easily further generalized, as it is straightforward to add additional exponential functions with further fitting parameters $a_1,a_2,a_3,...$. Because computations on exponential functions are relatively fast, these additional functions do not increase the computation time very much (also see next section). Of course, one could generalize any of the other ans\"atze by introducing a fitting parameter, but the increase in computation time is more problematic for these other ans\"atze.

\begin{figure*} [ht]
		\includegraphics[scale=.50]{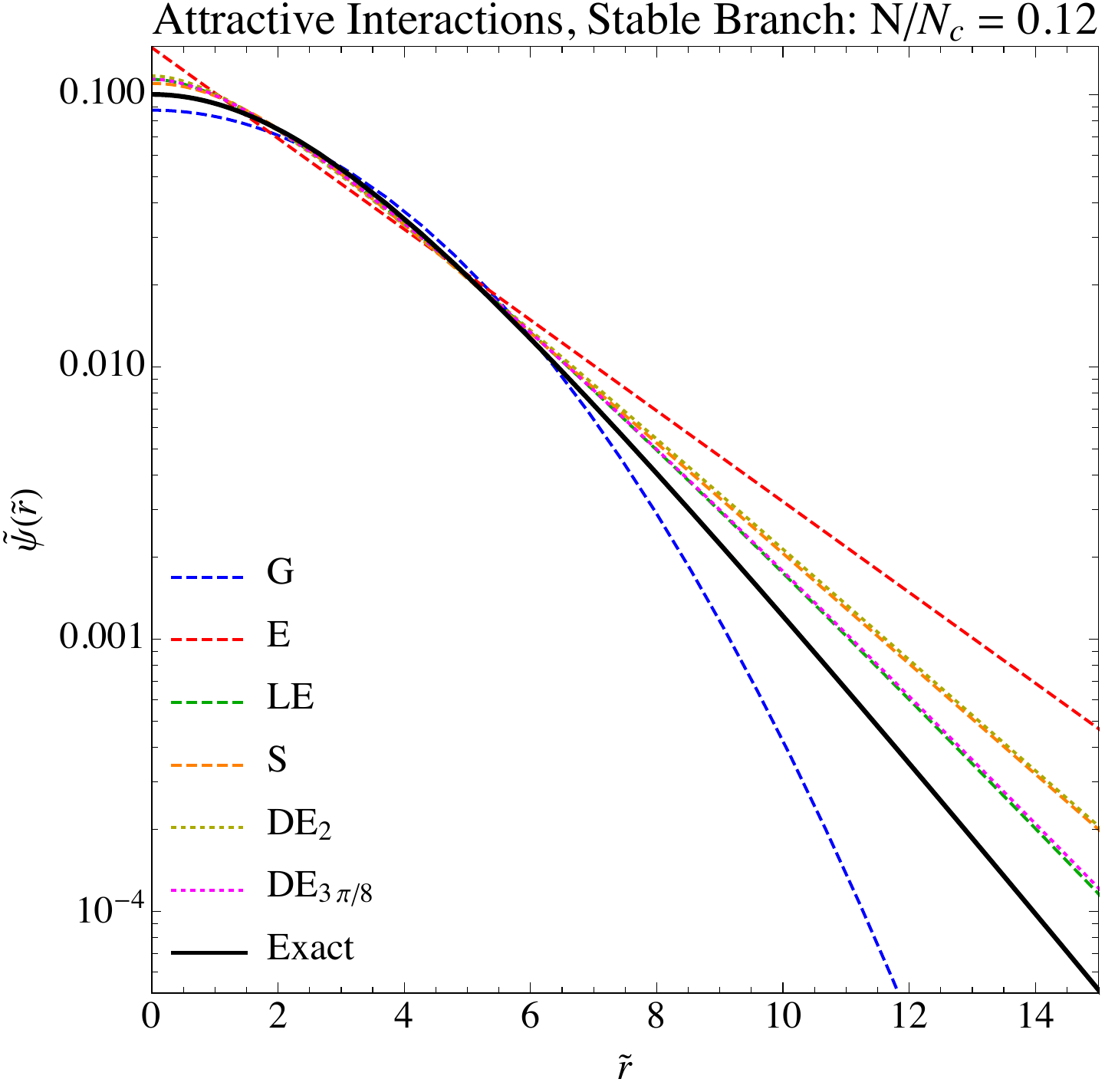}
	~ \quad
	 	\includegraphics[scale=.50]{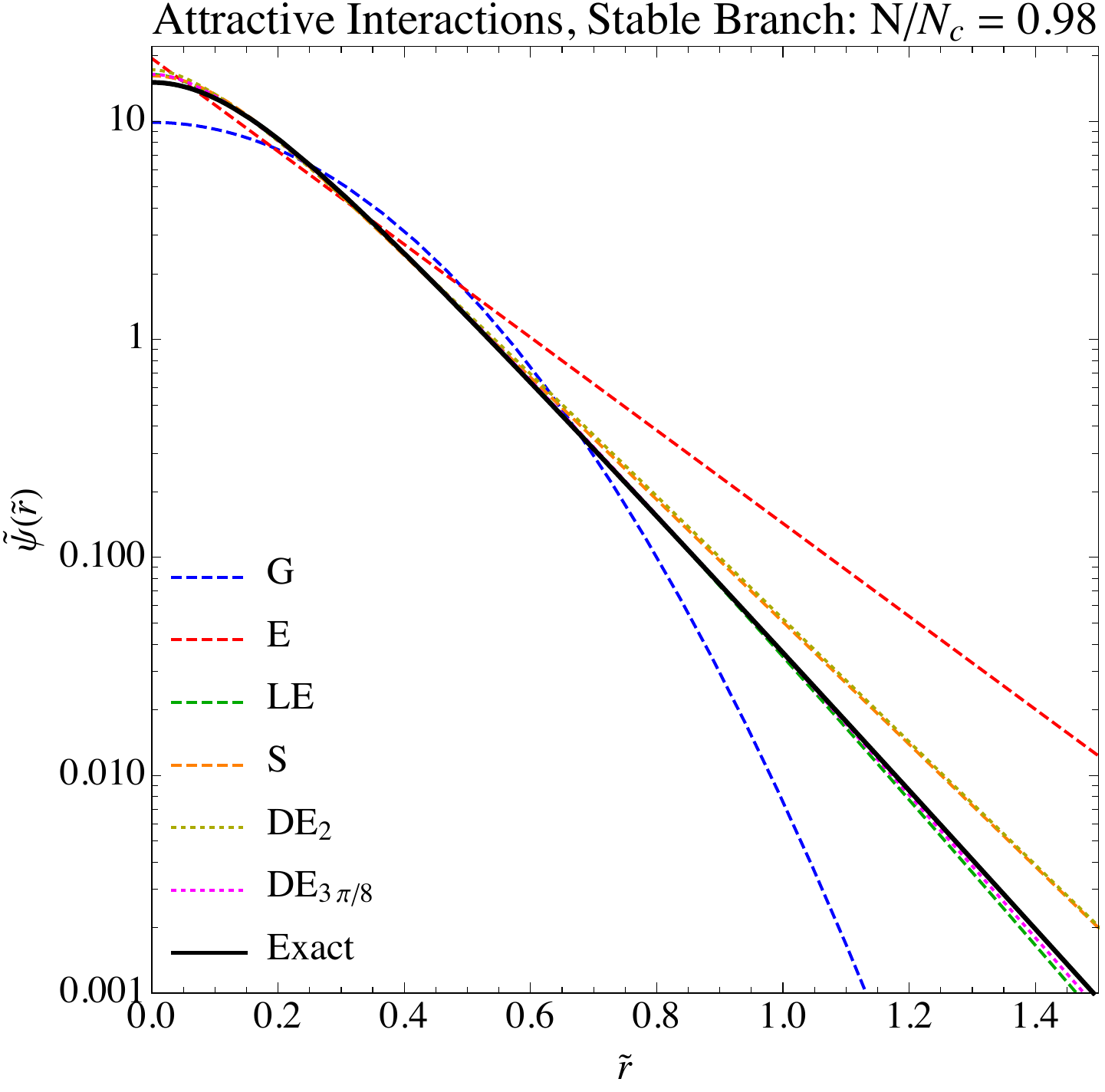}

		 \includegraphics[scale=.50]{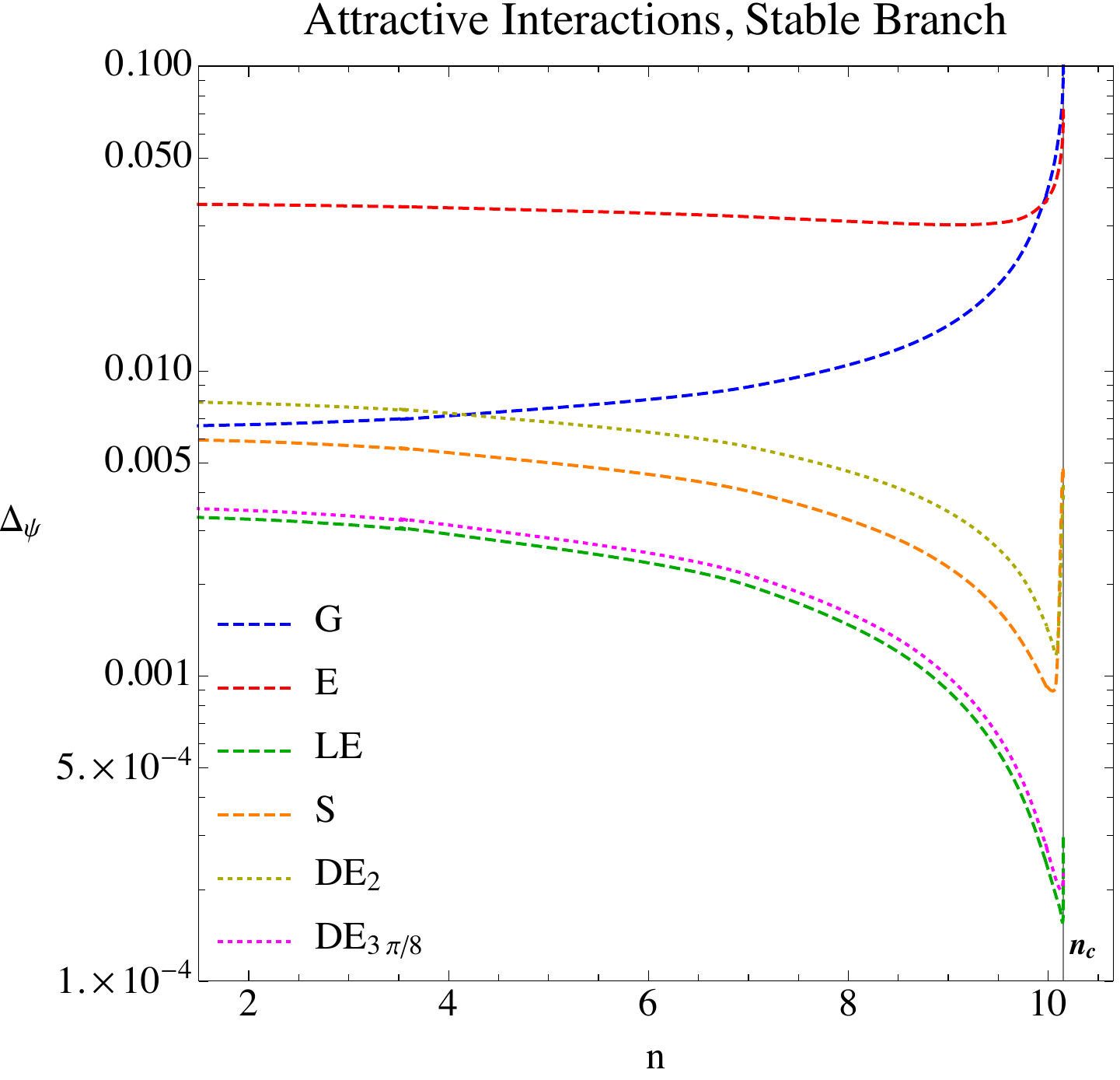}
	~ \quad
	 	 \includegraphics[scale=.48]{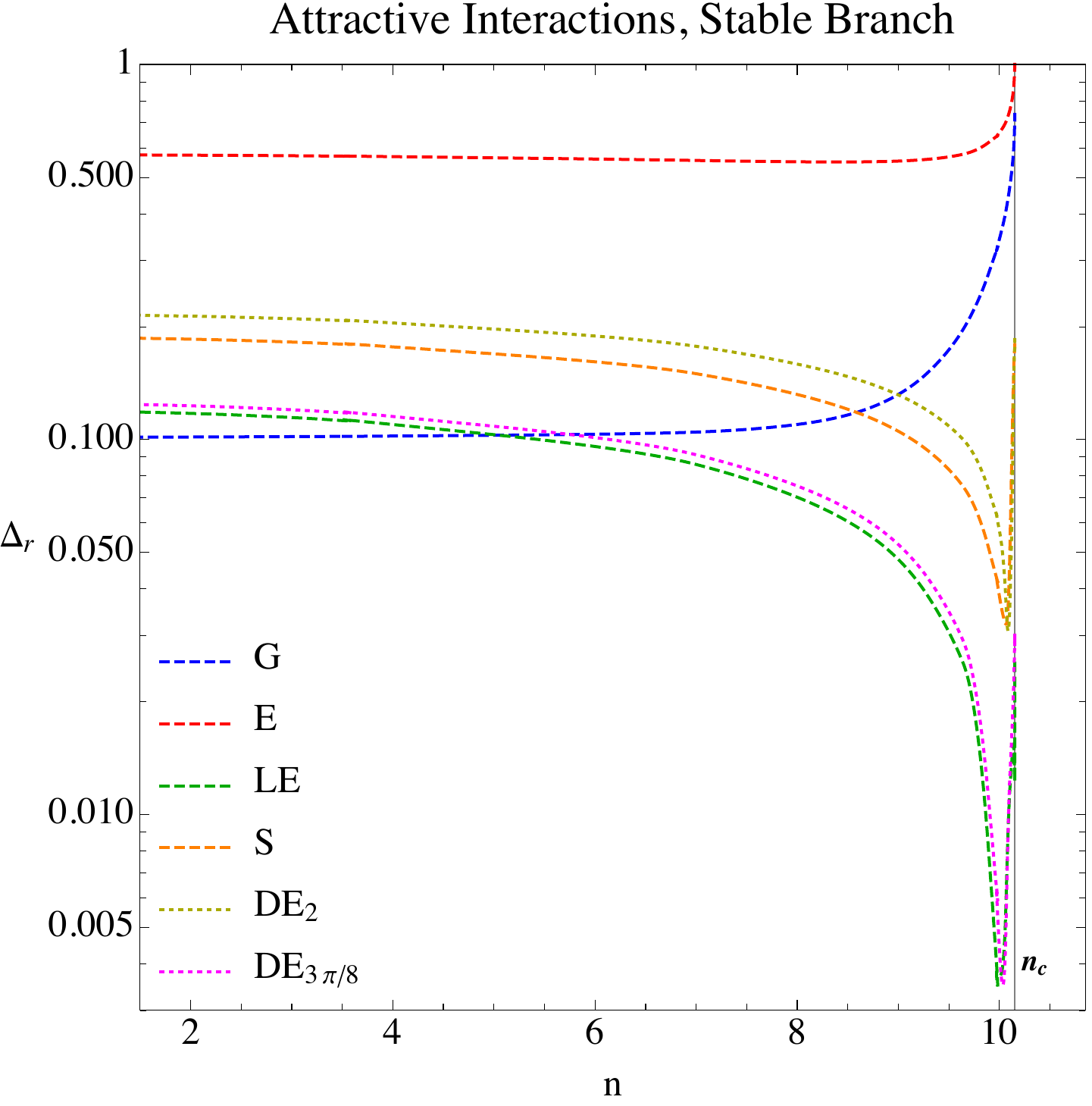}
	 \caption{Quantitative comparisons for attractive interactions, on the stable branch of solutions. In the upper two panels, we show the rescaled wavefunctions for two choices of particle number $N = 0.12N_c$ (left) and $N=0.98N_c$ (right). In the lower two panels, we show the corresponding deviations $\Delta_\psi$ (left) and $\Delta_r$ (right) of the ans\"atze from the exact solution, as defined in eqs. (\ref{eq:Deltar}) and (\ref{eq:Deltapsi}).}
 \label{fig:StableComp}
\end{figure*}

\subsection{Methods for Quantitative Comparisons} \label{sec:methods}

In this section, we report on the direct comparison of the exact numerical solutions of the previous section with the ans\"atze we presented in Section \ref{sec:Ansatze}. One criterion on which one can compare different ans\"atze is computational efficiency; in our analysis we have seen, for example, that the sech ansatz is significantly more difficult to employ than a function with exponential dependence. This is an important consideration because the purpose of using an ansatz in the first place is to simplify the calculation. The sech ansatz, for example, is orders of magnitude slower than the others (and, as we will show below, it does not pay off sufficiently in numerical accuracy). The double exponential is relatively fast, regardless of the value of $a$.

\begin{table}[t]
\centering
\hspace{-15pt}
\begin{tabular}{| c || c | c | c | c | c  |}
\hline
& Gaussian 
& Exp. 
& Lin $+$ Exp 
& Sech 
& Double Exp. \\
\hline \hline
Time [sec] & $2.28$ 
		& $2.48$ 
		& $4.67$
		& $358.2$
		& $\sim 11$
		\\ \hline
$\displaystyle{\frac{\text{Time}}{\text{Time}_{G}}}$ & $1$ 
		& $1.09$ 
		& $2.05$
		& $157.4$
		& $\sim 4.8$
		\\ \hline
  \hline
\end{tabular}
\caption{The computation time for the parameters in Table \ref{tab1}, for each ansatz. The top row are the absolute times in seconds on a standard laptop; the lower row is the time normalized to the Gaussian case.}
\label{tab_times}
\end{table}

\begin{figure*} [ht]
		\includegraphics[scale=.50]{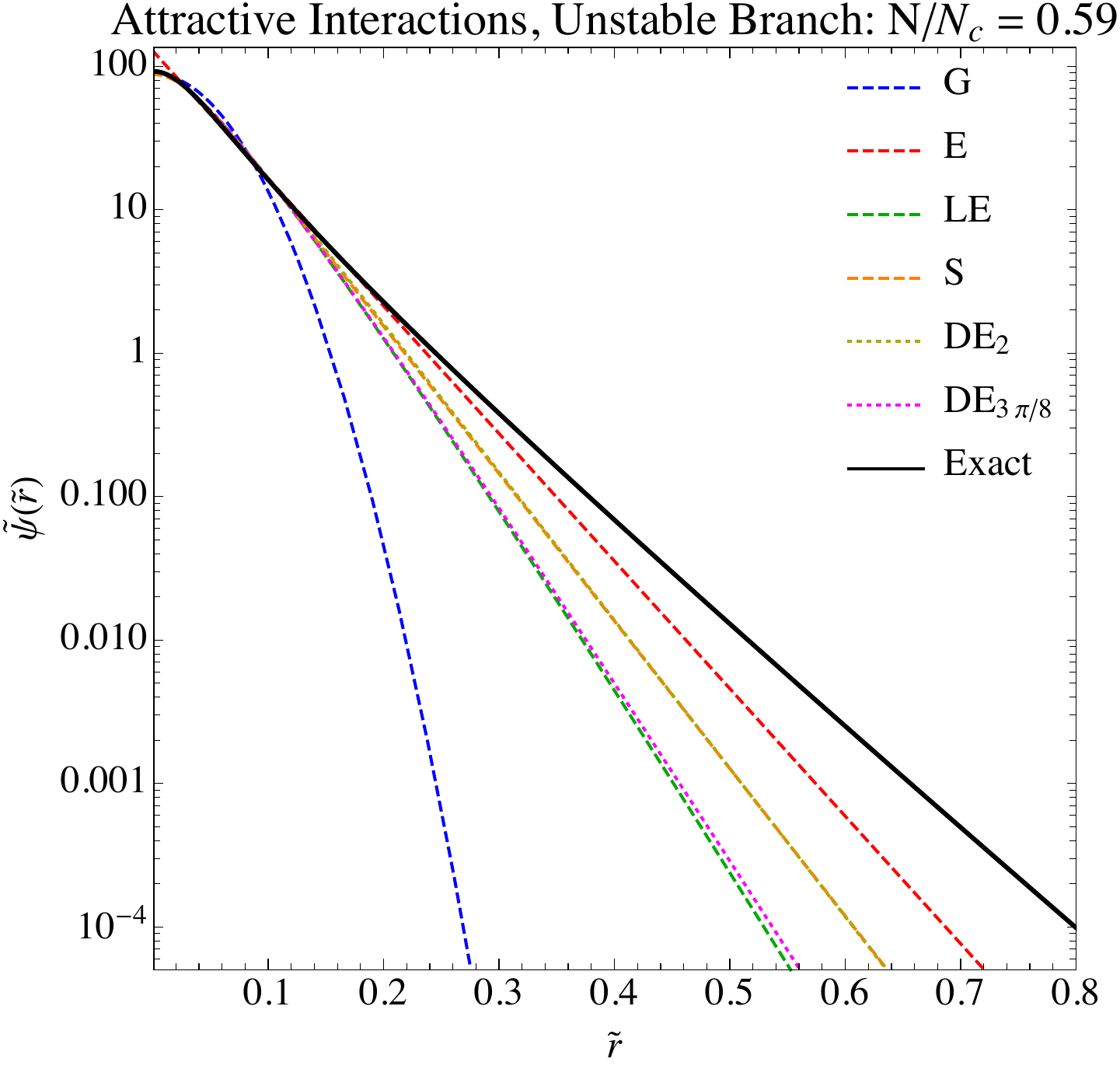}
	~ \quad
	 	\includegraphics[scale=.50]{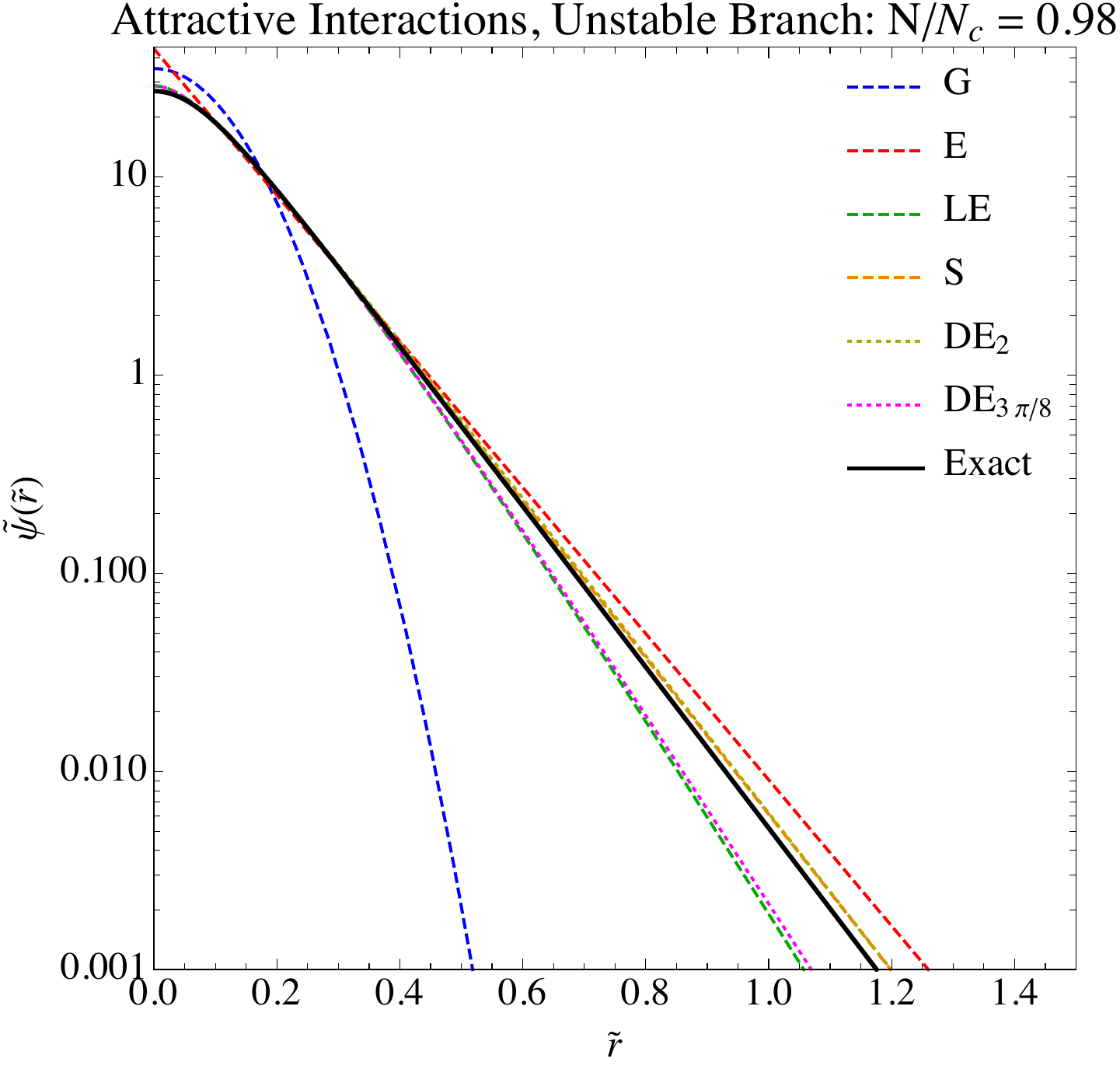}

		 \includegraphics[scale=.50]{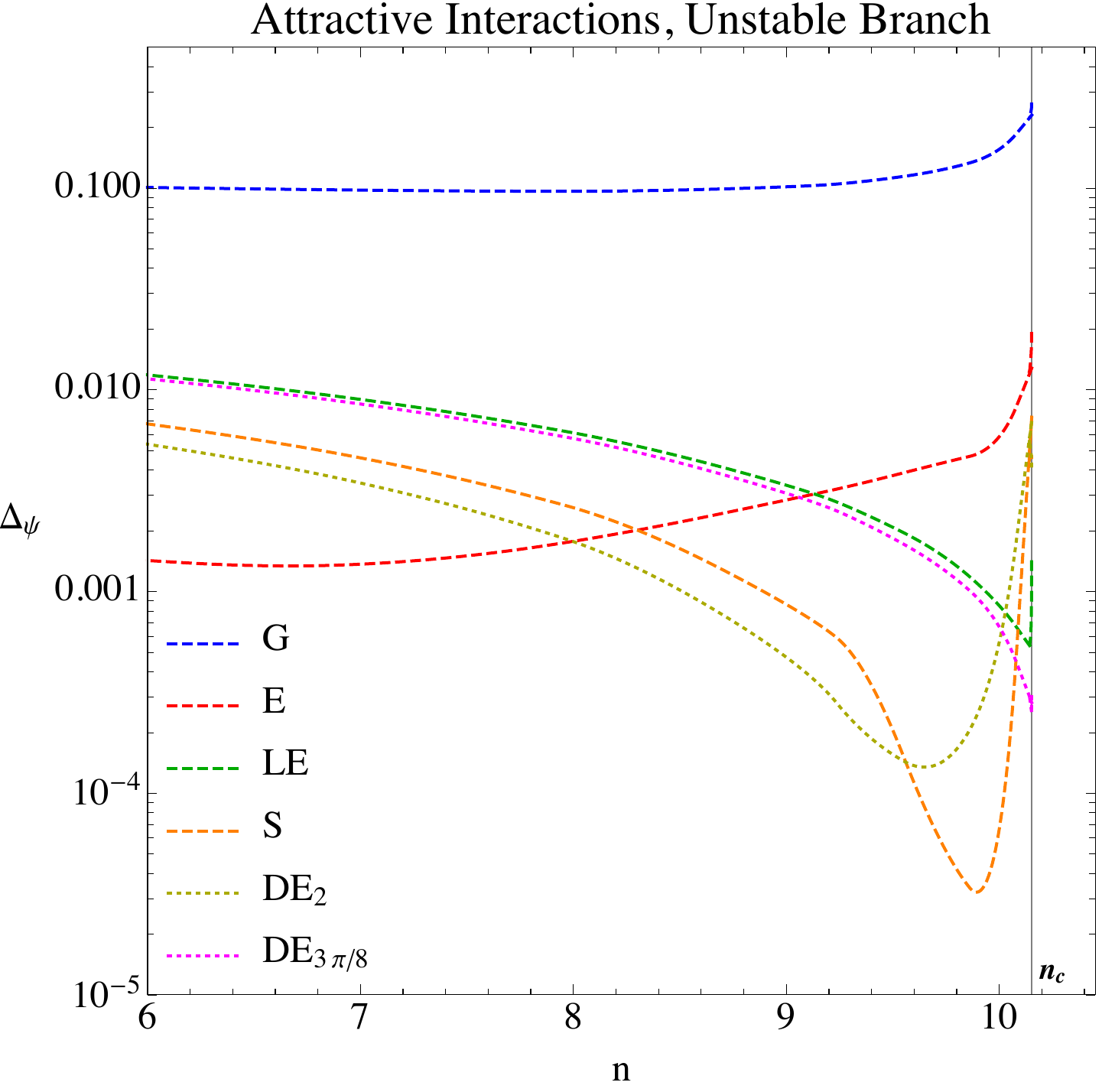}
	~ \quad
	 	 \includegraphics[scale=.48]{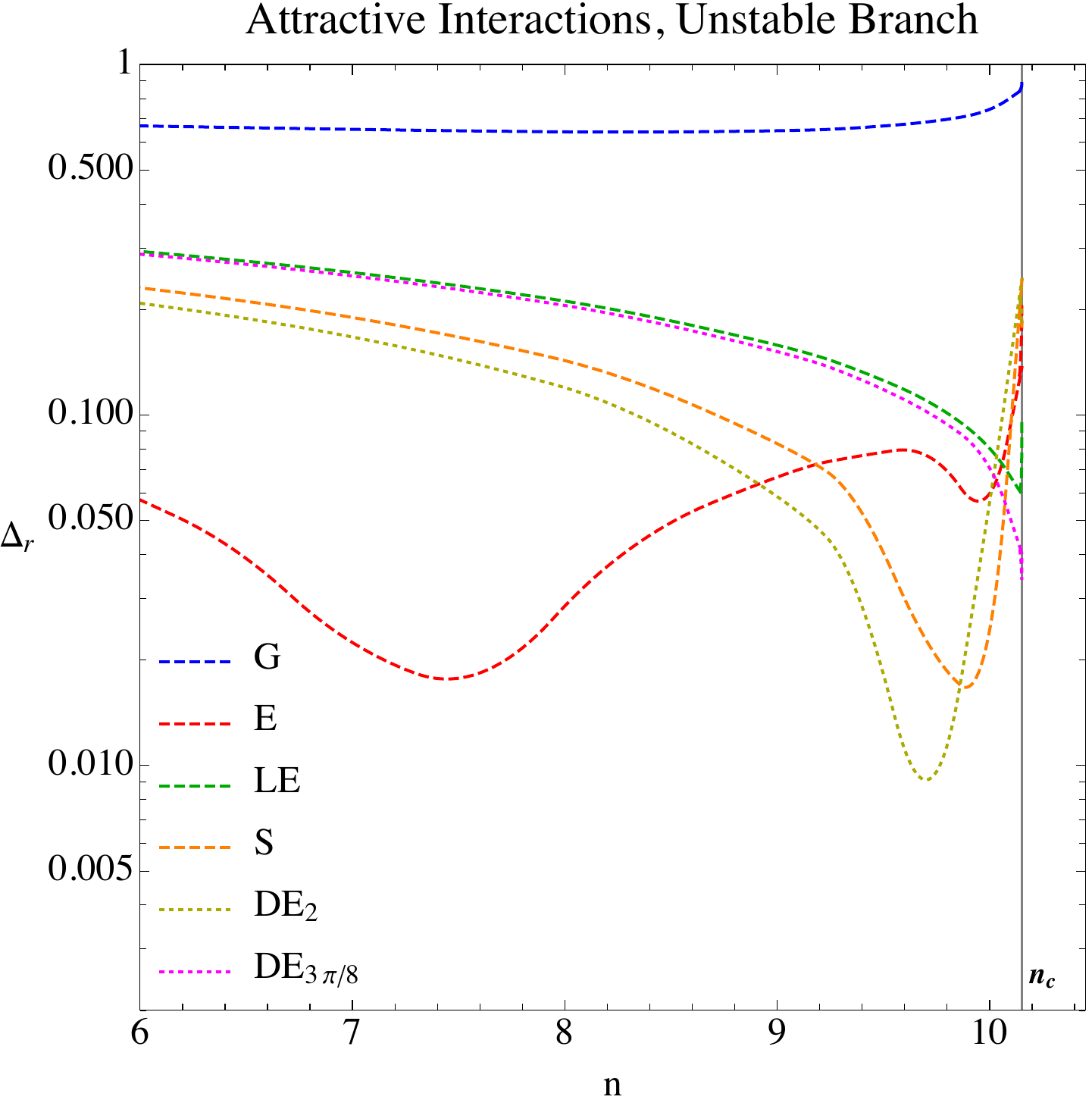}
	 \caption{Quantitative comparisons for attractive interactions, on the unstable branch of solutions. In the upper two panels, we show the rescaled wavefunctions for two choices of particle number $N = 0.59N_c$ (left) and $N=0.98N_c$ (right). In the lower two panels, we show the corresponding deviations $\Delta_\psi$ (left) and $\Delta_r$ (right) of the ans\"atze from the exact solution, as defined in eqs. (\ref{eq:Deltar}) and (\ref{eq:Deltapsi}).}
 \label{fig:UnstableComp}
\end{figure*}

Another criterion which is potentially more important is proximity to the exact solution. To make such comparisons simple to understand, we will rewrite the ans\"atze slightly so that the radial variable is the same as the numerical one, given in eq. (\ref{eq:numscale}); that is, we write
\begin{equation}
 \psi_A(\tilde{r}) = \frac{m^{5/2}}{M_P\,|\l|}\sqrt{\frac{n}{\rho^3\,C_2}} F_A\left(\frac{\tilde{r}}{\rho}\right),
\end{equation}
where $A$ labels the ansatz being considered, as before. Then
\begin{equation}
 \tilde{\psi}_A(\tilde{r}) \equiv \frac{M_P\,|\l|}{m^{5/2}}\psi_A(\tilde{r}) 
 			= \sqrt{\frac{n}{\rho^3\,C_2}} F_A\left(\frac{\tilde{r}}{\rho}\right)
\end{equation}
can be compared directly to the numerical result $\tilde{\psi}$ as a function of $\tilde{r}$.

A simple comparison of the shapes of the wavefunctions can be performed as follows: for a given $N$, compute $\vev{r}$, $\vev{r^2}$, and $R_{99}$ for the ansatz and compare it to the exact value computed numerically. These three radial variables should be a reasonable proxy for the shape of the wavefunction. More precisely, we will compute
\begin{equation} \label{eq:Deltar}
 \D_r(n) \equiv \sqrt{\left[\frac{\vev{r^2} - \vev{r^2}_{A}}{\vev{r^2}}\right]^2 +
 			\left[\frac{\vev{r} - \vev{r}_{A}}{\vev{r}}\right]^2 +
			\left[\frac{R_{99} - R^{A}_{99}}{R_{99}}\right]^2}.
\end{equation}

A potentially more robust method of comparison is to evaluate the numerical solution and a given ansatz at the same value of $N$, and see how large the wavefunction deviations are over the whole range of $r$. To make this quantitative, we compute the difference integral
\begin{equation} \label{eq:Deltapsi}
 \D_{\psi}(n) \equiv \frac{\int d^3\tilde{r} \left[\tilde{\psi}(\tilde{r}) - \tilde{\psi}_A(\tilde{r})\right]^2}
 					{\int d^3\tilde{r}\,\tilde{\psi}(\tilde{r})^2}.
\end{equation}
We will use both $\D_{\psi}$ and $\D_{r}$ to compare how well a given ansatz matches the exact result. In the following sections, we will show the result of these analyses for attractive interactions on both the stable and unstable branches, as well as for repulsive interactions.

\subsection{Attractive Interactions: Comparison for Stable Branch}

We begin with an analysis of the stable branch of solutions for attractive self-interactions. This branch is relevant for so-called dilute boson stars, including axion stars. In the top row of Figure \ref{fig:StableComp}, we show the exact and approximate wavefunctions, with the vertical axis on a log scale.  The top row correpsonds to attractive interactions along the stable branch. The most commonly used ans\"atze are the exponential and Gaussian functions, which turn out to be the worst fits to the numerical solutions, and are particularly bad in the large-$r$ tail. The other ans\"atze do reasonably well both near the core and in the tail.

Near the maximum mass, both the Gaussian and the exponential ans\"atze approximate the exact solution exceptionally badly. This is in part clear from Figure \ref{fig:MassRadius}: because these functions do not well-approximate the exact value of the maximum mass, the radius will also be very different, resulting in larger overall deviations (we will quantify these statements in a more precise analysis below). The other ans\"atze provide a much better fit in this region of parameter space. At large $\tilde{r}$, the double exponential does slightly better than the linear+exponential function.

For a more quantitative comparison, we turn to the bottom row of Figure \ref{fig:StableComp}. At every $n$, the smallest $\D_\psi$ deviations are found using the LE ansatz. In the limit $a\to1$, the same is found for the double exponential function, as discussed in an earlier section. In fact, we have scanned over values of the parameter $a$ and found that the best agreement to numerical results (i.e. smallest $\D_\psi$ and $\D_r$) is obtained in this limit. In this sense, the linear+exponential function provides the best fit to the data, and is also computationally efficient. It is also the most similar to the leading-order wavefunction of \cite{Kling1,Kling2}, whose wavefunction is explicitly calculated to have the correct asymptotic behavior as $r\to\infty$.

\begin{figure*} [ht]
		\includegraphics[scale=.50]{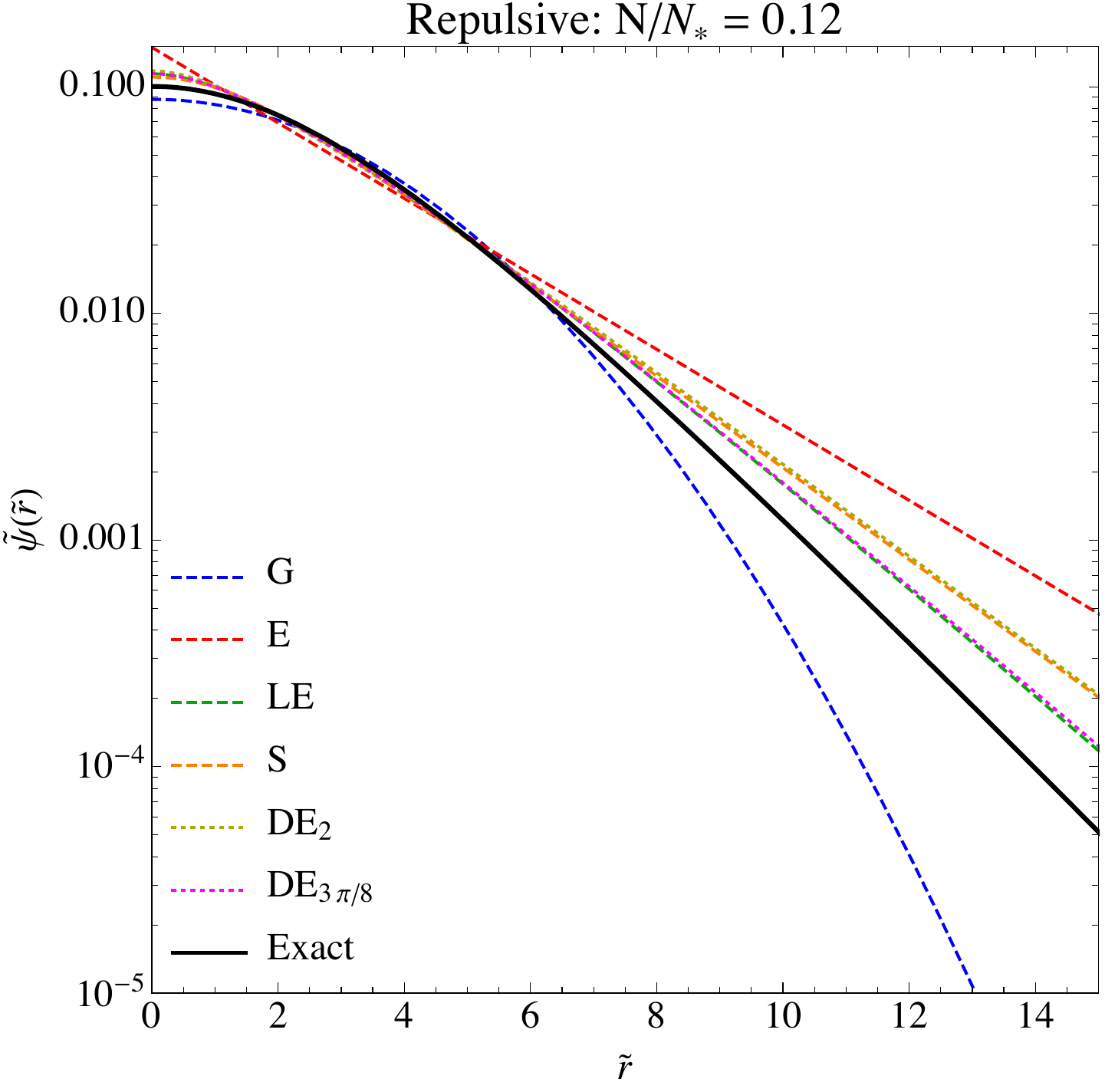}
	~ \quad
	 	\includegraphics[scale=.50]{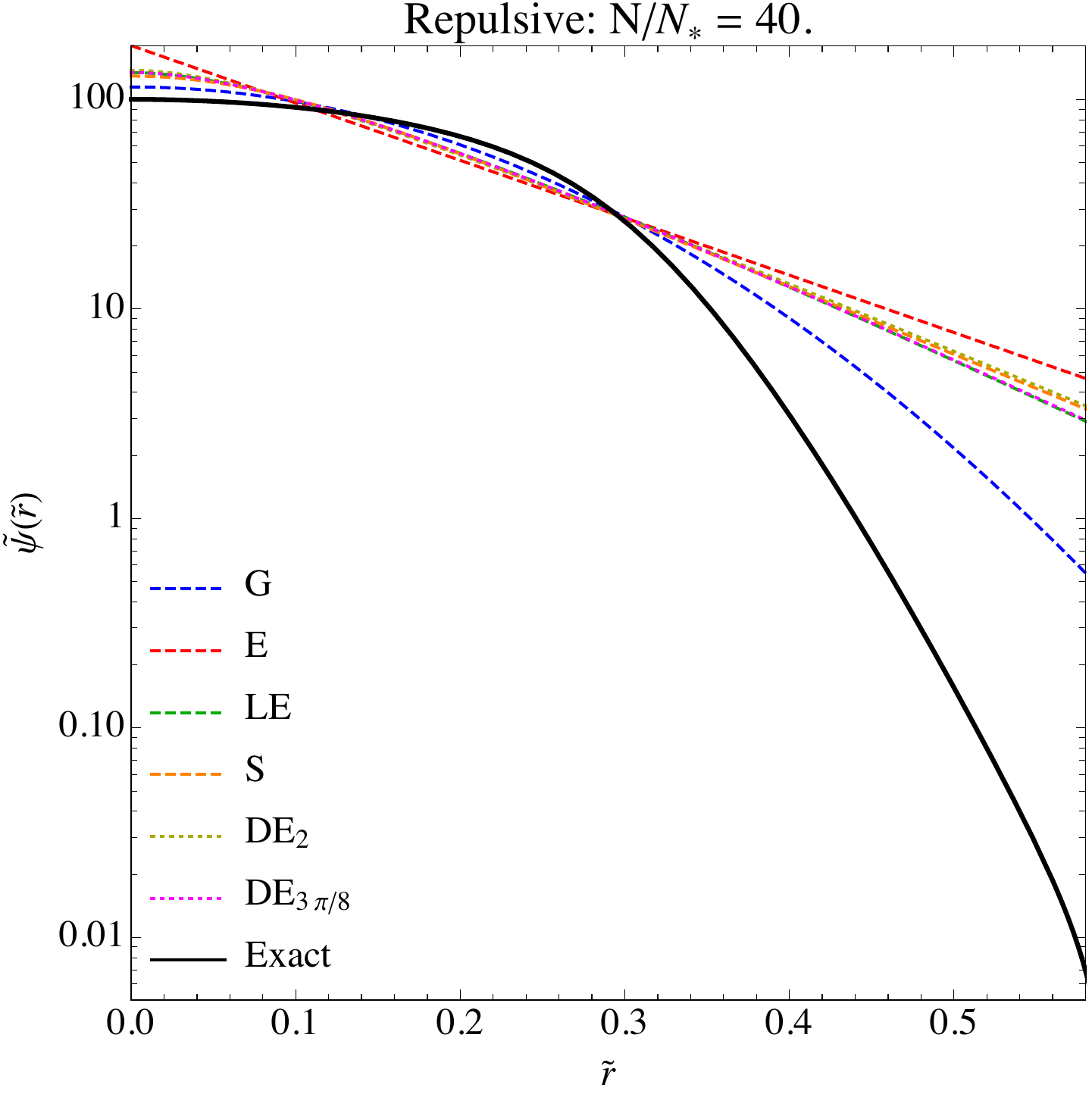}

		 \includegraphics[scale=.50]{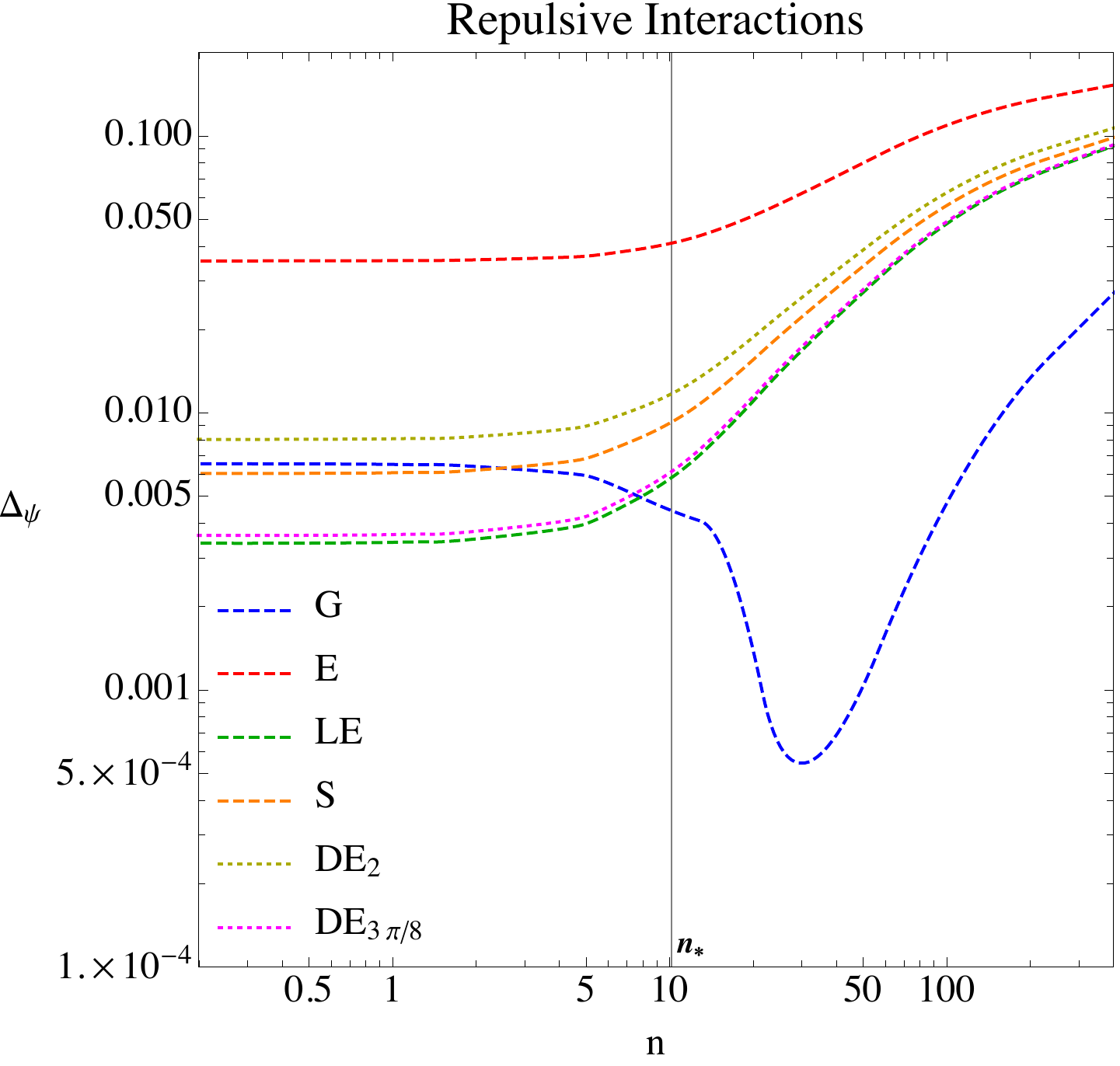}
	~ \quad
	 	 \includegraphics[scale=.48]{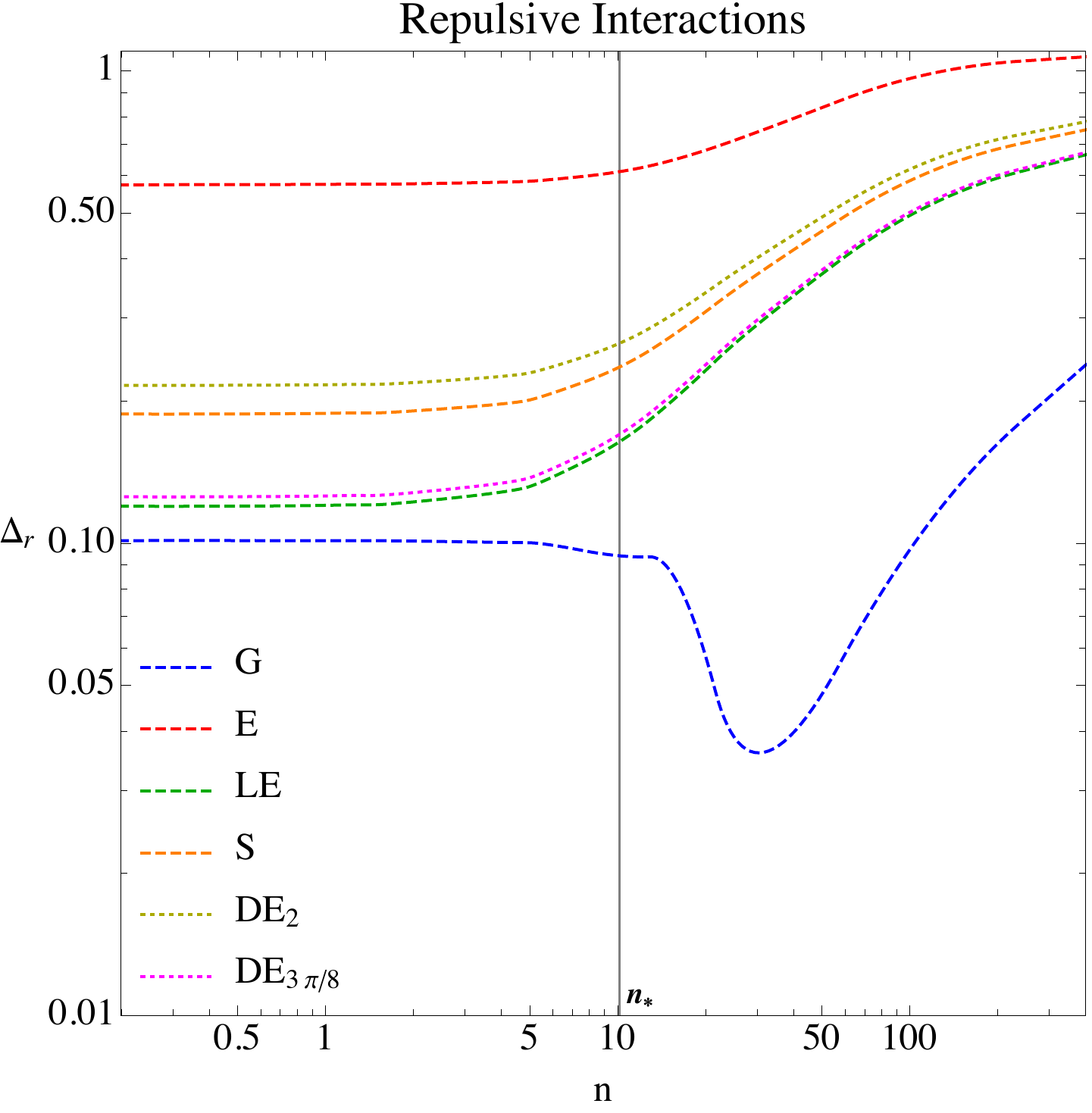}
	 \caption{Quantitative comparisons for repulsive interactions. In the upper two panels, we show the rescaled wavefunctions for two choices of particle number $N = 0.12N_*$ (left) and $N=40N_*$ (right). In the lower two panels, we show the corresponding deviations $\Delta_\psi$ (left) and $\Delta_r$ (right) of the ans\"atze from the exact solution, as defined in eqs. (\ref{eq:Deltar}) and (\ref{eq:Deltapsi}).}
 \label{fig:RepComp}
\end{figure*}

\subsection{Attractive Interactions: Comparison for Unstable Branch}

We move now to an analysis of the unstable branch which is relevant for collapsing boson stars, an important topic both for ordinary QCD axion stars \cite{ELSW,ELSW2}, as well as possible galaxy core collapse induced by astrophysical interactions \cite{ELSWFDM}. In the top row of Figure \ref{fig:UnstableComp}, we show the exact and approximate wavefunctions, with the vertical axis on a log scale for attractive interactions along the unstable branch.  It can be seen that for large $\tilde{r}$, $\psi_G$ is a poor fit for both small and large masses, while $\psi_S$ and $\psi_{DE_2}$ tend to be better fits than $\psi_{LE}$ and $\psi_{DE_{3\pi/8}}$.  It is also interesting to note that $\psi_E$ does better than all ans\"atze for some range of $n$, though we do not have a satisfactory explanation for this behavior at this time. We do observe in the numerical solutions that the wavefunctions become increasingly ``squeezed" towards $r=0$ on the unstable branch, and while the wavefunctions remain cored, they more closely approach an exponential behavior closer to the origin.

In the bottom row of Figure \ref{fig:UnstableComp}, we show the $\D_\psi$ and $\D_r$ deviations for the unstable branch.  For a range of masses, $\psi_E$ does better than all ans\"atze, while $\psi_S$ and $\psi_{DE_2}$ do better than $\psi_{LE}$ and $\psi_{DE_{3\pi/8}}$ for all masses except near the maximum mass. We conclude that over the most relevant parameter space, an exponential function is the ideal choice for analysis on the unstable branch, though close to $n_c$ it is preferable to use the double exponential with $a=2$. We do not recommend use of the sech function, only because of the computational inefficiency pointed out in Section \ref{sec:methods}.

\subsection{Repulsive Interactions: Comparison} \label{RepComp}

Finally, we move to an analysis of repulsive interactions which is relevant for axion-like particles; for recent model-building of dark matter scalar with repulsive interactions, see \cite{Fan}.  For such self-interactions, a maximum mass arises due to relativistic effects, to which our nonrelativistic analysis is not sensitive.  This maximum mass \cite{CSW}
\begin{equation}
M_{\text{max}}^{\text{rep}} = 0.22 \sqrt{\frac{\lambda}{4\pi}}\frac{M_P{}^3}{m^2}
\end{equation}
 is much greater than the masses resulting from our numerical solutions, and so the comparisons made here are all for physical masses.

For repulsive interactions, as the mass of the condensate increases, it approaches the Thomas-Fermi limit at which its kinetic energy becomes negligible and its radius becomes independent of mass \cite{DGPS_TF}. From eqs. (\ref{rhodilute}) and (\ref{nCrit}), one can see that as the number of particles $n$ increases, the solution for the equilibrium radius does, indeed, approach a constant.  This constant radius happens to be the critical radius for attractive self-interactions (eq. \ref{rhoCrit}), given that the ansatz considered is held constant.  It is also interesting to note that $N=N_*$ can be seen as a scale at which the condensate begins to approach the Thomas-Fermi limit.

In the top row of Figure \ref{fig:RepComp}, we show the exact and approximate wavefunctions, with the vertical axis on a log scale.  One can see that for $n \gg n_*$ all ans\"atze become increasingly worse fits in the tail. As is the case for attractive self-interactions, $\psi_{LE}$ and $\psi_{DE_{3\pi/8}}$ do better than $\psi_S$ and $\psi_{DE_2}$.  It is interesting to note that for $n/n_* \sim 40$, the wavefunction of the numerical solution falls off more rapidly at large $\tilde{r}$ than $\psi_G$.  This is in agreement with the results of B\"ohmer and Harko \cite{BH_TF} who found an exact solution to the equations of motion for the Thomas-Fermi limit to be of the form $\sqrt{\sin(\xi)/\xi}$.  This exact solution has been included as a possible compact ans\"atze to be used for axion-like condensates with repulsive interactions and large numbers of particles (see Section \ref{compact_ansatze}). For $N\ll N_c$, the condensate approaches the noninteracting limit; this is why the top-left panel of Figure \ref{fig:RepComp} is extremely similar to the top-left panel of Figure \ref{fig:StableComp}.

In the bottom row of Figure \ref{fig:RepComp}, we show the $\D_\psi$ and $\D_r$ deviations for repulsive interactions from which one can see that $\psi_G$ does better than all other ans\"atze at large $n$. It is also interesting to note that the deviations have a considerable change in behavior near $n_*$ (most notably for $\psi_G$), which is presumably due to the condensate approaching the Thomas-Fermi limit. For repulsive interactions at large $n$, the Gaussian function is thus the appropriate choice.

\section{Relativistic Formulation} \label{sec:RB}

In the preceding sections, we have analyzed various approximations to the GPP formalism. However, even the exact GPP equations are themselves an approximation to the underlying relativistic theory. The theory at high energies is defined by the Klein-Gordon equation for a scalar field, which is coupled to a curved spacetime metric described by the Einstein equations. In this section, we describe precisely the limit of the Einstein+Klein-Gordon (EKG) equations in which one recovers the GPP equations. We will then apply the relativistic formulation to the case of repulsive interactions, to analyze the limitations of the TF approximation.

\subsection{Equivalence between GPP and EKG Equations}

A relativistic method of describing boson stars was pioneered by Ruffini and Bonazzola (RB) \cite{RB}. The crucial idea is to take the field $\Acal$ describing the boson star to be linear in ground-state creation and annihilation operators $a_0^{(\dagger)}$,
\begin{equation} \label{fieldRB}
 \Acal(r,t) = R(r) \left[e^{-i\,\mu_0\,t}\,a_0 +
			  e^{i\,\mu_0\,t}\,a_0^\dagger\right],
\end{equation}
and use these operators to build $N$-particle states defined by
\begin{equation}
 \ket{N} = \frac{(a_0^\dagger)^N}{\sqrt{N!}} \ket{0}.
\end{equation}
Here, $\mu_0$ is the energy eigenvalue of the \emph{relativistic} equations; it is related to $\mu$ in the GPP formulation by $\mu = -(m-\mu_0)$. 
The proper normalization of the axion wavefunction $R(r)$  is \cite{RB}
\begin{equation}
 2\,\frac{\mu_0}{m}\,\int d^3r\,R(r)^2\,\sqrt{\frac{A(r)}{B(r)}} = 1,
\end{equation}
which effectively defines the particle number $N$.

The gravitational metric describing the curvature of spacetime in the presence of the boson star is
\begin{equation}
 ds^2 = -B(r)\,dt^2 + A(r)\,dr^2 + d\O^2,
\end{equation}
which is spherically symmetric, as is the ground state wavefunction $R(r)$ in eq. (\ref{fieldRB}). Then the $rr$ and $tt$ components of the Einstein equations (taken as expectation values), along with the Klein Gordon equation, constitute a complete system of equations for $R(r)$, $A(r)$, and $B(r)$:
\begin{align} \label{NEKGN}
 \bra{N}G_\m{}^\n\ket{N} &= 8\,\pi\,G\,\bra{N}T_\m{}^\n\ket{N} \nn \\
 KG[\Acal] = \Box\,\Acal - V'(\Acal) &= 0.
\end{align}
In the original RB paper, the self-interaction potential was trivial $V(\Acal)=m^2\Acal^2/2$, and only the $\bra{N}KG[\Acal]\ket{N-1}=0$ expecation value of the KG equation is nontrivial. In that case the ansatz of eq. (\ref{fieldRB}) is an exact solution. For the $\phi^4$ potential we have been considering, we have
\begin{equation}
 V(\Acal) = \frac{m^2}{2}\Acal^2 + \frac{\l}{4!}\Acal^4.
\end{equation}

Evaluating the expectation values as in \cite{RB}, we find\footnote{The third equation of (\ref{EKG1}) is the expectation value $\bra{N}KG[\Acal]\ket{N-1}=0$, but in the self-interacting theory higher-order expectation values $\bra{N}KG[\Acal]\ket{N-(2k+1)}=0$ with $k>0$ will not be satisfied, and so eq. (\ref{fieldRB}) is not an exact solution. However, as argued in \cite{GRB}, taking the leading order is a good approximation for all but the most strongly bound boson star configurations. In the context of the axion potential, we have presented a method of going beyond the RB ansatz to calculate relativistic corrections perturbatively in \cite{GRB}. For other work on relativistic corrections to scalar field theory, see \cite{GuthRelativistic,BraatenRelativistic,EMTWY}.}
\begin{align} \label{EKG1}
 &\frac{A'}{A^2\,r} + \frac{A-1}{A\,r^2}
	  = \frac{8\,\pi}{M_P{}^2}\Big[
		\frac{\mu_0{}^2\,N\,R^2}{B} + \frac{N\,R'^2}{A} \nn \\
	&\hspace{4cm}	 + N\,R^2 + \frac{N^2\,\l}{4\,m^2}R^4\Big],
		      \nn \\
 &\frac{B'}{A\,B\,r} - \frac{A-1}{A\,r^2}
	  = \frac{8\,\pi}{M_P{}^2}\Big[
		\frac{\mu_0{}^2\,N\,R^2}{B} + \frac{N\,R'^2}{A} \nn \\
	&\hspace{4cm}  - N\,R^2 - \frac{N^2\,\l}{4\,m^2}R^4\Big],
			\nn \\
 &R'' + \left(\frac{2}{r} + \frac{B'}{2\,B}
	  - \frac{A'}{2\,A}\right)R' \nn \\
	&\hspace{2cm} + A\left[\frac{\mu_0{}^2}{B}R
	  	-m^2\,R - \frac{N\,\l}{2}R^3\right]=0.
\end{align}
One can simplify the expression by defining \cite{ESVW}
\begin{equation}
 X(y) = \frac{2\,\sqrt{|\l|\,N}}{m}R(r), \qquad y = m\,r,
\end{equation}
in which case the EKG equations simplify to
\begin{align} \label{EKG2}
 &\frac{A'}{A^2\,y} + \frac{A-1}{A\,y^2}
	  = \frac{2\,\pi\,m^2}{M_P{}^2\,|\l|}\Big[
		\left(\frac{\mu_0}{m}\right)^2 \frac{X^2}{B} + \frac{X'^2}{A} \nn \\
	&\hspace{4cm}
	    + X^2  + \frac{\text{sgn}(\l)}{16}X^4\Big],
		      \nn \\
 &\frac{B'}{A\,B\,y} - \frac{A-1}{A\,y^2}
	  = \frac{2\,\pi\,m^2}{M_P{}^2\,|\l|}\Big[
		\left(\frac{\mu_0}{m}\right)^2\frac{X^2}{B} + \frac{X'^2}{A} \nn \\
	&\hspace{4cm}
	    - X^2 - \frac{\text{sgn}(\l)}{16}X^4\Big],
			\nn \\
 &X'' + \left(\frac{2}{y} + \frac{B'}{2\,B}
	  - \frac{A'}{2\,A}\right)X' \nn \\
	&\hspace{2cm}
	  + \frac{A}{2}\left[\left(\frac{\mu_0}{m}\right)^2\frac{X}{B}
	  	- X - \frac{\text{sgn}(\l)}{4}X^3\right]=0.
\end{align}
Note that primes in eq. (\ref{EKG2}) now indicate derivatives with respect to $y$ (rather than $r$).

In \cite{ESVW}, our original investigation of axion stars, we identified two small parameters and expanded the EKG equations in both. The analogue of these small parameters for a general theory with $\l\,\phi^4$ interactions is 
\begin{equation}
 \d = \frac{8\,\pi\,m^2}{M_P{}^2\,|\l|} \ll 1 \qquad \text{and} \qquad \D = \sqrt{1 - \frac{\mu_0{}^2}{m^2}} \ll 1.
\end{equation}
The requirement $\d\ll1$ corresponds to the weak gravity limit, as the $\d\to0$ limit recovers the Einstein equations for the vacuum. It is valid only when $|\l| \gg 8\,\pi\,m^2/M_P{}^2$. As an example, axions have $\l = -m^2/f^2$ and so this condition is equivalent to $8\,\pi\,f^2 \ll M_P{}^2$, a condition that is easily satisfied in nearly all applications. In particular, for QCD axions, $\d\approx10^{-14}$, and even theories of axionic ``fuzzy dark matter'', which have much larger $f$, still $\d\ll1$ is satisfied easily. The other parameter $\D$ is small precisely when the axion star is weakly bound, i.e. when the eigenenergy $\mu_0$ is of the same order as the particle mass $m$.

In applications with weak gravity, we can expand the metric components as
\begin{equation}
 A(r) = 1 + \d\,a(x), \qquad B(r) = 1 + \d\,b(x).
\end{equation} 
We also rescale the axion wavefunction $X(y)$ using
\begin{equation}
 Y(x) = \frac{1}{\D}\,X(y),
\end{equation}
with $x = y\,\D$ (which is to say, the wavefunction and coordinate scale with $\D$ as their scaling dimension). The resulting equations of motion, to leading order in $\d$ and $\D$, take the form
\begin{align} \label{EKG}
 a'(x) &= \frac{x}{2}\,Y(x)^2 - \frac{a(x)}{x}, \nn  \\
 b'(x) &= \frac{a(x)}{x}, \nn  \\
 Y''(x)&= - \frac{2}{x}Y'(x)-\frac{1}{8}\, Y(x)^3 
	  + [1 + \kappa\,b(x)]\,Y(x).
\end{align}
The constant $\k\equiv \d/\D^2$ controls the effective coupling to gravity, and is finite even though $\d,\D\ll1$. We have integrated this set of equations previously in \cite{ESVW} to find the spectrum of weakly bound axion stars.

Finally, we can integrate the first two equations to eliminate $a(x)$ and obtain a Poisson-like equation for $b(x)$:
\begin{equation}
 \nabla_x^2 b(x) = \frac{1}{2}\,Y(x)^2.
\end{equation}
This implies that $b(x)$ is proportional to the Newtonian gravitational potential $V_g$. The solution of this equation is
\begin{align}
 b(x) &= -\frac{1}{8\pi}\left[\int d^3 x' 
	\frac{Y(x')^2}{|\vec{x}-\vec{x'}|}\right].
\end{align}
Using this expression in the third equation in (\ref{EKG}), we arrive at a self-contained expression for the rescaled wavefunction $Y(x)$:
\begin{equation} \label{KGFinal}
 0 = \left[1 - \nabla_x^2 - \frac{1}{8}\,Y(x)^2
	-\frac{\kappa}{8\pi} \left( \int d^3 x' 
	\frac{Y(x')^2}{|\vec{x}-\vec{x'}|}\right)\right] Y(x).
\end{equation}
Because this equation is true only to leading order in $\D$, we refer to this equation as the \emph{infrared}, or low energy, limit of the Klein-Gordon equation for the axion.

We may manipulate the GPP equations (\ref{Poisson}) and (\ref{GPtime2}) into a very similar form in a straightforward way. First, rewriting the gravitational potential $V_g$ in integral form, we find a single integro-differential equation of motion
\begin{equation} \label{GP1}
 i\,\dot{\psi} = -\frac{1}{2m}\nabla^2 \psi 
      - \frac{\l}{8\,m^2}|\psi|^2\psi
      - G\,m^2\,\left[\int d^3r'\,\frac{|\psi(r')|^2}{|\vec{r}-\vec{r'}|}
	  \right]\,\psi.
\end{equation}
In the nonrelativistic limit, the wavefunction has a time dependence which is approximately harmonic, such that $i\dot{\psi} = \mu\,\psi = -(m-\mu_0)\psi \approx -m\,\D^2\,\psi / 2$. Further, we rescale to the dimensionless wavefunction
\begin{equation}
 \phi(r) = \sqrt{\frac{2\,|\l|}{m^3}}\frac{1}{\D} \psi
\end{equation}
and dimensionless coordinate $x = m\,\D\,r$ as before. Then eq. (\ref{GP1}) takes the form
\begin{align} \label{GPFinal}
 0 &= m^2\,\D^2\,\phi - m^2\,\D^2\,\nabla_x^2\phi 
	- \frac{m^2\,\Delta^2}{8}\phi^3 \nn \\
     & \hspace{4cm} - \frac{G\,m^4}{|\l|}\left[\int d^3x'\,
	    \frac{\phi(x')^2}{|\vec{x}-\vec{x'}|}\right]\phi  \nn \\
	&= m^2\,\D^2\left[1 - \nabla_x^2 - \frac{1}{8}\phi^2
      - \frac{G\,m^2}{|\l|\,\D^2}\left(\int d^3x'\,
	    \frac{\phi(x')^2}{|\vec{x}-\vec{x'}|}\right)\right]\phi.
\end{align}
Finally, we recognize the prefactor on the gravitational term as
\begin{equation}
  \frac{G\,m^2}{|\l|\,\D^2} = \frac{\delta}{8\pi\,\D^2} \equiv \frac{\kappa}{8\pi}.
\end{equation}
Thus, eq. (\ref{GPFinal}) is \emph{exactly equivalent} to eq. (\ref{KGFinal}) with the identification 
\[
 Y(x) \leftrightarrow \phi(x).
\]
Thus the GP formalism is equivalent to the RB formalism used in the infrared limit. We summarize the identifications between the two paradigms in Table \ref{tabEquiv}. 

\begin{table*}
\begin{center}
\begin{tabular}{| c | c | c |}
  \hline
  Name & GPP & Leading-Order RB \\
  \hline
  Equation(s) of Motion & Equation (\ref{GP1}) & Equations (\ref{EKG}) \\
  Wavefunction & $\displaystyle{\psi(r)}$ &
		  $\displaystyle{\sqrt{\frac{m}{2\,|\l|}}\,m\,\D\,Y(x)}$ \\
  Gravitational Potential & $\displaystyle{V_g(r)}$& 
		  $\displaystyle{\frac{4\,\pi\,G\,m^3}{|\l|}\,b(x)}$ \\
  Normalization & $\displaystyle{\int |\psi(r)|^2\,d^3r = N}$ & 
		  $\displaystyle{\frac{\D^2}{2|\l|}\int Y(x)^2\,d^3x = N}$\\
  \hline
\end{tabular}
\caption{Equivalence relations between the Gross-Pit\"aevskii+Poisson equations and Ruffini-Bonazzola equations governing axion stars in the low-energy limit. The coordinates are related by $x = m\,\D\,r$.}
\label{tabEquiv}
\end{center}
\end{table*}

Because these formalisms are precisely equivalent, one can work with whichever is more convenient for the application at hand. For example, in analyzing axion star decay, it is better to use the RB approach, as the transition matrix elements are more directly calculable \cite{ESW}. It is also more straightforward to generalize if higher-order relativistic corrections are needed \cite{GRB}. For the analysis of collapsing axion stars, we instead found it more convenient to use the GP formalism \cite{ELSW}.

\subsection{The Thomas-Fermi Approximation}

We will use the RB formulation to critically analyze the Thomas-Fermi (TF) approximation, which is commonly used in studies of repulsively interacting boson stars. At leading order, the equations of motion are
\begin{align} \label{eom}
 \nabla_y^2 b(y) &= \frac{1}{2}\,X(y)^2 \nn \\
 \nabla^2_y X(y) &- \Delta ^2\,X(y)-\delta \, b(y) \,X(y) - \frac{1}{8}\,X(y)^3=0,
\end{align}
where we have chosen to work with the wavefunction $X(y)$ rather than $Y(x)$ so that the appearance of the small parameters $\d,\D$ is manifest. In the TF approximation, one neglects the kinetic term $\nabla^2_y X(y)$ compared to the other terms. The consistency of that assumption must then be checked after the solution for $b(y)$ and $X(y)$ have been found.  Suppose the scale for $r$ is $R$, the radius of the boson star. Then the kinetic term is 
\begin{equation}
\nabla^2_y X(y) = \Ocal\left(\frac{X(y)}{m^2\,R^2}\right).
\end{equation}
Consequently, $\nabla_y^2 X$ is negligible compared to $\Delta^2\,X$ if 
\begin{equation}
R\gg \frac{1}{\Delta\,m}.
\end{equation}

If the TF approximation is valid, then the Klein-Gordon equation simplifies to
\begin{align}\label{eombZ}
 0 &= - \Delta ^2\,X(y)-\delta \, b(y) \,X(y) - \frac{1}{8}\,X(y)^3,
\end{align}
which can be directly solved for $X(y)$ to obtain
\begin{equation} \label{Xeq}
 X(y)^2 = -8\,\left(\d\,b(y) + \D^2\right).
\end{equation}
Substituting this back into the Poisson equation gives
\begin{equation}
 \nabla_y^2 b(y) = -4\,\left(\d\,b(y) + \D^2\right).
\end{equation}
The solution of this equation which is regular at the origin is
\begin{equation} \label{bsol}
 b(y) = -\frac{\D^2}{\d} - \frac{c}{y}\,\sin\left(p_0\,y\right),
\end{equation}
where $p_0 = 2\sqrt{\d}$ and $c$ is a dimensionless constant which will be determined below. Finally, we can calculate the wavefunction in the TF limit by substituting eq. (\ref{bsol}) into eq. (\ref{Xeq}):
\begin{equation} \label{XTFsol}
 X(y) = \sqrt{\frac{8\,\d\,c}{y}\,\sin\left(p_0\,y\right)}.
\end{equation}

We can now observe directly that the constant $p_0 = \pi / m\,R_{TF}$ defines the radius $R_{TF}$ of the condensate in the TF limit, given by
\begin{equation}
 R_{TF} = \frac{\pi}{m\,p_0} = \frac{\pi}{2m\sqrt{\d}} = \pi \sqrt{\frac{\l}{32\pi}}\frac{M_P}{m^2}
\end{equation}
and matches the standard result \cite{BH_TF}, previously derived by considering a polytropic equation of state. Note that the radius of the condensate in the TF limit does not depend on the particle number $N$; this behavior of the cutoff is characteristic of the TF approximation. The coefficient $c$ can be determined by normalization of the wavefunction (c.f. Table \ref{tabEquiv})
\begin{equation}
 N = \frac{1}{m} \int d^3r\,T_{00} \approx \frac{1}{2\,\l}\int d^3y \,X(y)^2 = \frac{16\pi^2\,c\,\d}{p_0{}^2\l} = \frac{4\pi^2\,c}{\l},
\end{equation}
which implies $c = N\,\l / (4\pi^2)$. Thus, we can write down the final expression for the rescaled wavefunction
\begin{equation}
 X(y) = \frac{2\,\sqrt{\d^{3/2}\,N\,\l}}{\pi} \sqrt{\frac{\sin(p_0\,y)}{p_0\,y}}.
\end{equation}

Finally, we consider the breakdown of the TF approximation. To check the self-consistency of omitting the kinetic term from the equation of motion, (\ref{eom}), we must substitute the solutions of eqs. (\ref{XTFsol}) and (\ref{bsol})  back into (\ref{eom}), to see whether the kinetic term is  much smaller than the rest of the terms. Substituting those solutions into the kinetic term gives a complicated expression, which simplifies considerably if we evaluate it just at the boundaries of the physical range for $r$, at $r=0$ and $r=R_{TF}=\pi\,/\,(m\,p_0).$ In both limits we obtain, up to numerical constants,
\begin{equation} \label{KinTF}
\nabla_y^2 X(y) \sim \d\,\sqrt{\d^{3/2}\,N\,\l}.
\end{equation}
Similarly, evaluating the interaction term at the the and of the range of $r$ we obtain
\begin{equation} \label{IntTF}
X(y)^3 \sim (\d^{3/2}\,N\,\l)^{3/2}.
\end{equation}
The validity of the Thomas-Fermi approximation requires that the ratio of eq. (\ref{KinTF}) and (\ref{IntTF}) be small, i.e.
\begin{equation}\label{constraint2}
 \frac{\d}{\d^{3/2}\,N\,\l} = \frac{1}{\sqrt{\d}\,N\,\l} \ll 1.
\end{equation}
This implies
\begin{equation}
 N \gg N_{\rm min,TF} \sim \frac{1}{\sqrt{\d}\,\l} \sim \frac{M_P}{m\,\sqrt{\l}},
\end{equation}
and comparing to eq. (\ref{scaling}) we see that this condition simply implies $n\gg 1$.

A further constraint on the Thomas-Fermi solution is the relationship between the Schwarzschild radius, $R_{\rm SCH}$ and $R$.  Unless $R>R_{\rm SCH}$ the boson star collapses to a black hole.  This constraint leads to the inequality
\begin{equation}
R_{TF} = \pi\,\sqrt{\frac{\l}{32\pi}} \frac{M_P}{m^2} \gg R_{\rm SCH}=2\,N\,\frac{m}{M_P{}^2}
\end{equation}
In other words, there is an upper limit for $N$,
\begin{equation} 
N \ll N_{\rm max,TF} = \sqrt{\frac{\pi\,\l}{128}} \frac{M_P{}^3}{m^3}.
\end{equation} 
Note that this differs from the exact result of Colpi et al. \cite{CSW} only by a small numerical factor.
Of course, if $N\ll N_{\rm max}$ is not satisfied, then our calculations, based on the Newtonian approximation to gravity, would not be acceptable. Therefore we obtain upper and lower (\ref{constraint2}) bounds for $N$ in the Thomas-Fermi approximation:
\begin{equation}\label{limits}
\frac{M_P}{m\sqrt{\lambda}} \ll N_{\rm TF} \ll \sqrt{\frac{\pi\,\l}{128}} \frac{M_P{}^3}{m^3},
\end{equation}
providing a wide range for the applicability of the approximation, provided 
\begin{equation}\label{constraint3}
\lambda\gg \frac {m^2}{M_P^2}.
\end{equation}
The TF approximation has been generalized to describe rotating \cite{Shiraishi1} or charged \cite{Shiraishi2} boson stars, as well as boson stars comprised of $N$ scalar fields \cite{Shiraishi3}.

Note that the behavior of the TF wavefunction near $R_{TF}$ is not physical at finite $N$. One could in principle calculate corrections to the TF equations of motion which take this into account, which should give rise to the standard exponentially falling wavefunction at large $r$. This would have consequences in certain applications where the tail behavior is important, including calculations of the classical decay rate. An analysis of this type is beyond the scope of the present paper.

\section{Conclusions} \label{sec:Conclusions}

In this analysis we have analyzed a number of approximate methods for describing boson stars. We focused on the Gross-Pit\"aevskii+Poisson (GPP) equations, which have a broad range of applicability for weakly-bound, nonrelativistic boson stars.
Using a time-independent variational formalism, we compared various ans\"atze which describe gravitationally bound BECs with self-interactions. These ans\"atze allow the GPP system to be solved analytically, alleviating the obligation to cumbersome numerical solutions. Moreover, numerical solutions exist primarily for stationary BEC configurations, and are much more difficult to use in dynamical applications; ans\"atze are powerful tools for solving dynamic BEC problems such as collapse, collisions, and expansion.

We have treated the numerical solution to the stationary GPP system as a benchmark for comparing different ans\"atze, prioritizing factors such as computational ease, fit of the wavefunction profile, and value of maximum mass (for attractive interactions). We found that a linear+exponential wavefunction is the best fit for attractive self-interactions along the stable branch, as well as for repulsive self-interactions at small $N$.  For attractive self-interactions along the unstable branch  a single exponential is the best fit for small $N$ while a sech wavefunction fits better for large $N$, though the latter is computationally inefficient.  A Gaussian wavefunction, which is used often in the literature, is exceedingly poor across most of the parameter space, with the exception of repulsive interactions for large $N$. 

We found that our proposed double-exponential ansatz is much more tunable compared to other ans\"atze in the literature. By choosing various values of the parameter $a$, one may optimize a given fit parameter over others, or near-perfectly replicate more computationally complex ans\"atze. In particular, on the stable branch for attractive interactions, the limit $a\to1$ gives rise to the linear+exponential ansatz, which we found to be in closest agreement to the exact case. On the unstable branch, one can take the $a\gg1$ limit to obtain the exponential ansatz, appropriate at large central densities, while the choice $a=2$ very nearly reproduces the sech ansatz, which is quite inefficient computationally.  We also showed how to generalize the addition of free parameters in order to create more computationally efficient ans\"atze using the double exponential. 
 
There remain many unstudied applications of the ans\"atze we have compiled here, and analytic solutions to the time-dependent GPP system will undoubtedly have myriad uses. As mentioned in the introduction, there are many open questions regarding oscillons formed by inflation and quintessence fields, and axion BECs remain the subject of active study.  Along with a rigorous, multi-criterion comparison of ans\"atze from the relevant literature, we have presented arguments for the necessity of time-dependent ans\"atze, and have introduced a tunable, computationally simple ansatz. 

Finally, we also highlighted the relevant differences between the relativistic Einstein+Klein-Gordon (EKG) equations, as derived using the Ruffini-Bonazzola formalism, and the nonrelativistic GPP system. In particular, we showed explicitly the relevant expansion parameters which characterize the nonrelativistic limit. We used this formulation to critically analyze the Thomas-Fermi approximation, which is a large-$N$ limit for repulsive self-interactions in which the kinetic energy is taken to be small.

\section*{Acknowledgements}
We thank N. Bar, F. Kling, M. Ma, B. Maddock, and C. Vaz, for discussions.  The work of J.E. was supported by the Zuckerman STEM Leadership Program. J.E. also thanks the Galileo Galilei Institute for Theoretical Physics for the hospitality and the INFN for partial support during the completion of this work. M.L. thanks the Barry Goldwater Scholarship and Excellence in Education Foundation for scholarship support. M.L. and L.S. thank the Department
of Physics at the University of Cincinnati for financial support in the form of Violet Diller Fellowships.

 \appendix

\section{Alternate Formulation for Boson Star Ans\"atze} \label{AppB}
 \renewcommand{\theequation}{A.\arabic{equation}}

An alternative formulation, used for example in \cite{ChavanisMR,EKNW,ESVW2}, is as follows: The energy per particle is
\begin{equation} \label{EperN}
\epsilon(\s) \equiv \frac{E}{N}
	= \frac{A}{\sigma^2}-\frac{B\,N}{\sigma} \pm \frac{N\,C}{\sigma^3},
\end{equation}
where by comparison with (\ref{binding}), we identify
\begin{equation}
A = \frac{1}{m}\,\frac{D_2}{2\,C_2}, \quad 
B = \frac{m^2}{M_P{}^2}\,\frac{B_4}{2\,C_2{}^2}, \quad 
C = \frac{|\l|}{m^3}\,\frac{C_4}{16\,C_2{}^2}.
\end{equation}
The energy per particle of (\ref{EperN}) is minimized at
\begin{equation}
\sigma_{sol}=\frac{A}{B\,N}\left[1 + \sqrt{1 + \text{sgn}(\l) \frac{3\,N^2\,B\,C}{A^2}}\right].
\end{equation}

\begin{table*}[ht]
\centering
\begin{tabular}{| c || c |}
\hline
& DE$_a$\\
\hline \hline
 $F(\xi)$ & $\dis{\frac{1}{a-1}\left(ae^{-\xi} - e^{-a \xi}\right)}$
		\\ \hline
 $B_4$ & \begin{tabular}{c}
 		$\dis{\frac{8 \pi ^2}{a\,(a-1)^4} \left[\frac{5 a^5}{64} + \frac{5}{64 a^4} - 4 a+\frac{10}{a+1}
				- \frac{6}{3a+1} + \frac{19}{2 (a+1)^2}  - \frac{270}{(a+3)^2} - \frac{6}{(3 a+1)^2} \right.}$ \\
		$\dis{\left.  - \frac{119}{4(a+1)^3} + \frac{324}{(a+3)^3} + \frac{4}{(3a+1)^3} + \frac{30}{(a+1)^4} 
				- \frac{10}{(a+1)^5} + \frac{49}{4}\right]}$
 		\end{tabular}
 		\\ \hline
 $C_2$ &  $\dis{\frac{4 \pi}{(a -1)^2}  \left(\frac{1}{4 a^3}+\frac{a ^2}{4}-\frac{4 a }{(a
   +1)^3}\right)}$
		\\ \hline
 $C_4$ & $\dis{\frac{4 \pi}{(a-1)^4}  \left(\frac{a ^4}{32}-\frac{8 a ^3}{(a +3)^3}+\frac{1}{32 a
   ^3}+\frac{3 a ^2}{2 (a +1)^3}-\frac{8 a }{(3 a +1)^3}\right)}$
		\\ \hline
 $D_2$ &  $\dis{\frac{\pi\,a^2}{(a-1)^2} \left(1+\frac{1}{a^3}-\frac{16}{(1+a)^3}\right)}$
		\\ \hline
  \hline
\end{tabular}
\caption{Table containing relevant numerical parameters for the double exponential ansatz; the rows correspond to the first few rows of Table \ref{tab1}. The other parameters, including $\bar{n}$ and $\rho_c$, can be calculated using these values and eqs. (\ref{nCrit}) and (\ref{rhoCrit}). The ratio $\k \equiv R_{99}/\s_{d}$, has no closed form for arbitrary values of $a$ (but is simple to calculate given a particular value).}
\label{tabDE}
\end{table*}

In the special case of $\l<0$ (attractive self-interactions), we find the critical number and radius beyond which no bound state solutions exist,
\begin{align}
N_c &= \frac{A}{\sqrt{3\,B\,C}}, \nn \\
\sigma_c &= \frac{A}{B\,N_c} = \sqrt{\frac{3\,C}{B}}.
\end{align}
We can also write the bound state solution in terms of $N_c$ and $\sigma_c$,
\begin{equation}
\sigma_{sol} = \sigma_c\left[1 + \sqrt{1 - \left(\frac{N}{N_c}\right)^2}\right]
		= \sigma_c \, (1 + \delta),
\end{equation}
where $\d \equiv \sqrt{1 - N^2/N_c{}^2}$. In these variables the energy per particle can then be written as
\begin{equation}
\epsilon(\sigma_{sol}) = \sqrt{\frac{B^3}{3\,C}} \left[
		\frac{N_c}{(1+\delta)^2} - \frac{N}{(1+\delta)}
		- \frac{N}{3}\frac{1}{(1+\delta)^3}\right].
\end{equation}
The corresponding value for the chemical potential is
\begin{equation}
\m(\sigma_{sol}) = \sqrt{\frac{B^3}{3\,C}} \left[
		\frac{N_c}{(1+\delta)^2} - \frac{2\,N}{(1+\delta)}
		- \frac{2\,N}{3}\frac{1}{(1+\delta)^3}\right].
\end{equation}
Evaluated at the critical values (i.e. $N=N_c$ and $\d=0$), the energy per particle and chemical potential are,
\begin{align}
\epsilon(\sigma_c) &= -\frac{A\,B}{9\,C}, \nn \\
\mu(\sigma_c) &= -\frac{5\,A\,B}{9\,C}.
\end{align}
We thus arrive at the very interesting result for the ratio
\begin{equation}
 \frac{\m(\s_c)}{\e(\s_c)} = 5,
\end{equation}
which will be true independently of the choice of ansatz.

\section{Full Solutions for Double Exponential Ansatz} \label{AppC}
Tabulated in Table \ref{tabDE} are the full solutions for the parameters in Table \ref{tab1} for the double exponential ansatz for arbitrary values of $a$.  Note that $\bar{n}$ and $\rho_c$ can be calculated from these coefficients by using eqs. (\ref{nCrit}) and (\ref{rhoCrit}).

\end{document}